\PassOptionsToPackage{table,dvipsnames}{xcolor}
\documentclass[manuscript,screen]{acmart}
\usepackage{amsmath} 
\usepackage{amsthm} 
\usepackage{graphicx} 
\usepackage{float} 
\usepackage{multirow} 
\usepackage{array}
\usepackage{enumitem}
\usepackage{subcaption}
\usepackage{colortbl}
\usepackage{booktabs}

\definecolor{shade0}{RGB}{255,255,255}
\definecolor{shade1}{RGB}{255,250,240}
\definecolor{shade2}{RGB}{255,245,225}
\definecolor{shade3}{RGB}{255,240,210}
\definecolor{shade4}{RGB}{255,230,190}
\definecolor{shade5}{RGB}{255,220,170}
\definecolor{shade6}{RGB}{255,210,150}
\definecolor{shade7}{RGB}{255,200,130}
\definecolor{shade8}{RGB}{255,190,110}
\definecolor{shade9}{RGB}{255,180,90}

\newcommand{\heatmanual}[2]{\cellcolor{shade#1} #2}

\newcounter{exboxctr}
\renewcommand{\theexboxctr}{\arabic{exboxctr}}
\newsavebox{\ExampleBox}

\newenvironment{Example}[1]{%
  \refstepcounter{exboxctr}%
  \par\vspace{1em}
  \setlength{\fboxsep}{5pt}%
  \setlength{\fboxrule}{0.4pt}%
  \begin{lrbox}{\ExampleBox}%
    \begin{minipage}{0.85\linewidth}
      \textbf{Example \theexboxctr:} #1
      \par\smallskip
      \hrule height 0.8pt 
      \par\smallskip
}{%
    \end{minipage}
  \end{lrbox}%
  \begin{center}
    \fbox{\usebox{\ExampleBox}}%
  \end{center}
  \par\smallskip
}

\newcolumntype{L}[1]{>{\raggedright\arraybackslash}p{#1}}
\newcolumntype{C}[1]{>{\centering\arraybackslash}p{#1}}

\AtBeginDocument{%
  }

\setcopyright{acmlicensed}
\copyrightyear{2025}
\acmYear{2025}
\acmDOI{XXXXXXX.XXXXXXX}

\acmConference[Conference acronym 'XX]{Make sure to enter the correct
  conference title from your rights confirmation emai}{June 03--05,
  2018}{Woodstock, NY}
\acmISBN{978-1-4503-XXXX-X/18/06}

\begin{document}

\title{Evaluating the Robustness of Retrieval-Augmented Generation to Adversarial Evidence in the Health Domain}

\author{Shakiba Amirshahi}
\orcid{0009-0009-1524-8643}
\affiliation{%
  \institution{University of Waterloo}
  \city{Waterloo}
  \state{ON}
  \country{Canada}
}
\email{shakiba.amirshahi@uwaterloo.ca}

\author{Amin Bigdeli}
\orcid{0009-0003-8977-9312}
\affiliation{%
  \institution{University of Waterloo}
  \city{Waterloo}
  \state{ON}
  \country{Canada}
}
\email{abigdeli@uwaterloo.ca}

\author{Charles L. A. Clarke}
\orcid{0000-0001-8178-9194}
\affiliation{%
  \institution{University of Waterloo}
  \city{Waterloo}
  \state{ON}
  \country{Canada}
}
\email{claclark@gmail.com}

\author{Amira Ghenai}
\orcid{0000-0001-5583-3008}
\affiliation{%
  \institution{Toronto Metropolitan University}
  \city{Toronto}
  \state{ON}
  \country{Canada}
}
\email{aghenai@torontomu.ca}

\renewcommand{\shortauthors}{Amirshahi et al.}

\begin{abstract}

Retrieval augmented generation (RAG) systems provide a method for factually grounding the responses of a Large Language Model (LLM) by providing retrieved evidence, or context, as support. Guided by this context, RAG systems can reduce hallucinations and expand the ability of LLMs to accurately answer questions outside the scope of their training data.
Unfortunately, this design introduces a critical vulnerability: LLMs may absorb and reproduce misinformation present in retrieved evidence.
This problem is magnified if retrieved evidence contains adversarial material explicitly intended to promulgate misinformation. This paper presents a systematic evaluation of RAG robustness in the health domain and examines alignment between model outputs and ground-truth answers.
We focus on the health domain due to the potential for harm caused by incorrect responses, as well as the availability of evidence-based ground truth for many common health-related questions. We conduct controlled experiments using common health questions, varying both the type and composition of the retrieved documents (helpful, harmful, and adversarial) as well as the framing of the question by the user (consistent, neutral, and inconsistent). Our findings reveal that adversarial documents substantially degrade alignment, but robustness can be preserved when helpful evidence is also present in the retrieval pool. These findings offer actionable insights for designing safer RAG systems in high-stakes domains by highlighting the need for retrieval safeguards.
To enable reproducibility and facilitate future research, all experimental results are publicly available in our github repository\footnote{\url{https://github.com/shakibaam/RAG_ROBUSTNESS_EVAL}}.

\end{abstract}




\maketitle

\section{Introduction}
\label{sec:introduction}
 
Large Language Models (LLMs) have become central to modern information access systems, powering a broad range of applications in fields such as question answering \cite{Liu:2023:PPP,arefeen2024leancontext,naveed2025comprehensive}. However, their deployment introduces new risks to information reliability, particularly due to their susceptibility to hallucinations and the generation of factually incorrect responses \cite{bender2021dangers,onoe2022entity,zhang2023language,huang2025survey,chang2024survey}. To resolve these issues and improve factual grounding, many such systems adopt a Retrieval-Augmented Generation (RAG) architecture, which combines the generalization capabilities of LLMs with document-level evidence retrieved from an external corpus to enhance the credibility and accuracy of output answer~\cite{Xiong:2024:BRG,Gao:2024:RGL}.
When answering questions, RAG systems employ retrieved documents as context to provide factual ground for their responses.

While the architecture of RAG systems is designed to enhance accuracy and trustworthiness, it also introduces new vulnerabilities as the reliability of LLM output becomes contingent on the integrity of the retrieved documents employed as context. Therefore, if malicious or misleading content is injected into the retrieval corpus, whether unintentionally or through deliberate manipulation by attackers, LLMs may confidently generate harmful misinformation. 

Under this threat model, the retrieval system and its neural ranking model (NRM) act as an attack vector for adversarial documents, which are designed to exploit scoring and ranking functions, thereby elevating documents containing malicious content into the top-k ranks~\cite{wu2023prada,liu2023black,chen2023towards,bigdeli2024empra}. When supplied as context to the LLM, these documents then serve as a payload, delivering manipulated content that can directly influence the generation process. This two-stage vulnerability mirrors classical security architectures, where an exploit (vector) delivers a harmful effect (payload) to a dependent system.

A growing body of literature demonstrates that LLMs are susceptible to such adversarial attacks~\cite{zou2402poisonedrag,chaudhari2024phantom,nazary2025poison,zhang2024hijackrag,BadRAG}. For example, \citet{zhang2024hijackrag} craft adversarial documents by appending the user query with a fixed instruction prompt designed to elicit a specific answer from the LLM. In a related study, \citet{chaudhari2024phantom} concatenates the user query with templated instruction prompts to achieve various attack objectives, such as inducing refusals to answer or generating biased responses. \citet{zou2402poisonedrag} adopt a similar strategy by pairing the user query with an LLM-generated passage that embeds a malicious answer.

Despite demonstrating the feasibility of adversarial manipulation in RAG pipelines, existing attack strategies face several notable limitations and challenges: (1) Attacks that append the query to manipulated documents, commonly referred to as Query+ in the IR adversarial literature, are easily detectable, as they introduce repetitive patterns that can be flagged by both human evaluators and automated filters designed to detect semantic redundancy~\cite{bigdeli2024empra,chen2023towards,wu2023prada}. (2) Fixed-format prompt injections or simplistic LLM-generated texts lacking rhetorical diversity, which could be detected by safety systems similar to appending the query to the document. (3) These attacks do not evaluate LLM vulnerabilities on high-stakes topics where models are explicitly trained to resist misinformation, such as health domains.

Our work does not introduce a new attack method but instead provides an empirical robustness evaluation of LLMs under controlled conditions, where real and realistic documents containing misinformation are introduced as RAG context. To the best of our knowledge, no prior study has empirically tested RAG robustness in medically critical domains using realistic human-annotated and adversarially constructed documents, making this the first systematic assessment in such a high-risk setting. Specifically, we investigate the vulnerabilities of LLMs in both RAG and Non-RAG settings in the health domain, especially their sensitivity to health-related misinformation, such as counterfactual claims about COVID-19, where incorrect or misleading results can have severe real-world consequences. We focus on precisely those domains where models are expected to be most resilient due to targeted alignment and safety training, thereby providing a more stringent and realistic robustness assessment. Unlike prior work that relies on synthetic, templated, or low-stakes content, we employ naturalistic, high-risk adversarial documents designed to evade pre-processing and semantic similarity filters. Specifically, we: (1) evaluate the susceptibility of RAG systems to adversarial documents in the medical domain where these documents are provided as retrieval context for medically consequential queries, and (2) analyze how presuppositions and tone embedded in user queries interact with retrieved content to bias the LLM’s interpretation, agreement, and final response.

To this end, we conduct experiments using test collections from the TREC 2020 and TREC 2021 Health Misinformation Tracks \cite{DBLP:conf/trec/ClarkeMS21,DBLP:conf/trec/ClarkeRSMZ20}, which provide medically focused queries accompanied by gold-standard human-written documents labeled as helpful, harmful, or neutral by expert assessors based on relevance, correctness, and credibility. Conventional large-scale benchmarks such as MS MARCO \cite{nguyen2016ms} or the TREC Web tracks \cite{TREC2020}, as well as other existing datasets, are unsuitable for our purposes, as they provide only topical relevance judgments and do not include per-query annotations distinguishing helpful from harmful documents or accounting for factual accuracy and misinformation. To complement these human-annotated corpora, we also employ LLM-generated adversarial documents released by \citet{bigdeli2025fsap}. An adversarial document is defined as a document that has been deliberately manipulated or generated to appear credible while introducing misleading or incorrect information, with the explicit goal of disrupting retrieval performance. In their work, \citet{bigdeli2025fsap} generated such documents for each helpful item in both TREC 2020 and TREC 2021 using a range of adversarial generation strategies. These include methods that invert factual claims, reframe content to introduce subtle inaccuracies, and inject misleading evidence, thereby producing realistic but harmful alternatives that can evade detection, ranking above authentic, helpful documents.

To evaluate the interaction between document content and user query framing, we construct three distinct versions of each query to reflect different user presuppositions: (i) a \textit{consistent} version aligned with the query’s ground-truth stance, (ii) a \textit{neutral} version, representing the original query from the test collection, and (iii) an \textit{inconsistent} version that contradicts the ground-truth stance. By pairing these with helpful, harmful, and adversarial documents, we evaluate LLMs in both RAG (Ragnarok framework~\cite{Pradeep:2025:RRF}) and Non-RAG settings. Importantly, we assume documents are already retrieved, which allows us to isolate the generation stage from the retrieval process. This enables a focused evaluation of LLM behavior under controlled document-query conditions.

Based on this setup, our study addresses the following research questions (RQs):

\begin{itemize}
    \item \textbf{RQ1:} Does retrieved context influence LLM responses? 
    
    \item \textbf{RQ2:} To what extent does the type of retrieved context (helpful, harmful, and adversarial) influence robustness? 
    
     \item \textbf{RQ3:} In what ways do orders and combinations of multiple evidence sources shape model behavior? 
    
    \item \textbf{RQ4:} How do different query framings (consistent, neutral, and inconsistent) influence ground-truth alignment? 
\end{itemize}

Our contribution lies in providing a systematic empirical characterization of these vulnerabilities and the contextual factors that influence their severity. Although the susceptibility of RAG systems to adversarial documents has been established conceptually, the quantitative impact across different document types, their interaction with query framing, and their effectiveness against safety-aligned models in the high-stakes health domain remains underexplored. By quantifying the degree to which adversarial documents undermine ground truth alignment, and by analyzing how query framing interacts with retrieved content, our study provides an evidence base for understanding and mitigating risks. Through systematic measurement across query variations and document combinations, we establish baseline vulnerability metrics and provide the empirical foundation necessary for developing robust retrieval architectures. This systematic evaluation addresses a critical gap between theoretical attack feasibility and practical risk assessment, transforming abstract security concerns into measurable risks for evidence-based system design.

More concretely, the contributions of our work include:
\begin{itemize}
    \item We introduce a controlled evaluation framework for measuring RAG robustness against adversarial exposure in high-stakes health domains.
    
    \item We systematically analyze how query framing and document type jointly influence ground-truth alignment, offering insights into how implied user presuppositions interact with retrieved evidence.

    \item We provide empirical findings showing that while adversarial documents substantially degrade ground-truth alignment, LLMs exhibit greater resilience to COVID-related misinformation compared to general health misinformation, suggesting that post-training alignment influences domain-specific robustness.
\end{itemize}

\section{Related Work}

\subsection{RAG Systems}
RAG systems emerged in part as an approach to overcoming the hallucination problem of LLMs, and to increase accuracy in generated responses by combining their generative capabilities with factual grounding provided by external knowledge~\cite{lewis2020retrieval,Gao:2024:RGL,guu2020retrieval}. In a basic RAG pipeline, a retrieval component first identifies and ranks relevant documents from a large corpus. Then, the top-k documents are passed to an LLM as {\em context} for generating the final response. This RAG architecture has been widely adopted for question answering and similar applications~\cite{siriwardhana2023improving,Liu:2023:PPP,arefeen2024leancontext}.

Recent research has increasingly centered on evaluating and improving the performance of RAG systems, while also introducing tailored architectures and task-specific adaptations \cite{gao2024smartrag,liu2023lost,cuconasu2024power,zhou2022docprompting,zhang2024multi,izacard2020leveraging,izacard2022few,macdonald2025constructing,sudhi2024rag,wang2024feb4rag}. For instance, \cite{liu2023lost} explored the impact of the positioning of the documents within the context window on the output generated by the LLM. In a similar study, \citet{cuconasu2024power} investigated the impact of the positioning of not only relevant documents but also noisy irrelevant documents on the RAG system. \citet{gao2024smartrag} introduced a RAG pipeline that jointly optimizes its policy network, retriever, and answer generator via reinforcement learning, achieving better performance and lower retrieval cost than separately trained modules. While these works provide valuable insights into RAG behavior, they primarily focus on the RAG settings and their performance. In contrast, our work investigates RAG systems under adversarial conditions, where the context may be intentionally crafted to mislead the LLM, thereby exposing vulnerabilities at the intersection of retrieval and generation.

\subsection{Prompt Framing in LLMs and RAG Systems}

Recent advancements in LLMs have demonstrated impressive performance across various natural language processing tasks.
However, their sensitivity to prompt formulation remains a notable challenge~\cite{qiang2024prompt,razavi2025benchmarking}. Minor modifications in the wording, structure, or even punctuation of prompts can cause outputs that are substantially different and often incorrect~\cite{qiang2024prompt,razavi2025benchmarking,mao2023prompt,wang2024assessing,mizrahi2024state,li2025enhancing,perccin2025investigating,hu2024prompt}.
\citet{qiang2024prompt} demonstrated that fine-tuned LLMs suffer large performance drops when prompts are perturbed by synonyms or paraphrases, introducing Prompt Perturbation Consistency Learning to enforce prediction stability across such variants. In related work, ~\citet{mao2023prompt} conducted a comprehensive study of prompt position and demonstrated that the location of prompt tokens in input text has a large effect on zero-shot and few-shot performance, with many widely used placements proving suboptimal.

In the RAG setting, prompt effects also remain critical. ~\citet{perccin2025investigating}  investigated the robustness of RAG pipelines at the query level, illustrating that subtle prompt variations, such as redundant phrasing and shifts in formal tone, can considerably affect retrieval quality and overall accuracy. Complementing these findings, ~\citet{hu2024prompt} analyzed prompt perturbations in RAG and proposed the Gradient Guided Prompt Perturbation, an adversarial method that steers model outputs toward incorrect answers, while also presenting a detection approach based on neuron activation patterns to mitigate these risks.

In this paper, we extend this line of work by examining how different query framings interact with retrieval context in RAG systems focused on the health domain. We evaluate consistent, neutral, and inconsistent query styles to measure their influence on ground-truth alignment and robustness under adversarial retrieval conditions. This design captures the ways in which query framing amplifies or mitigates the effects of helpful and harmful evidence, offering a systematic view of prompt sensitivity in a high-stakes domain.

\subsection{Adversarial Attacks on Neural Retrieval Models and LLMs}

A basic RAG system comprises two primary components: a retriever, which retrieves and ranks relevant documents from an external corpus, and an LLM, which generates responses based on the retrieved context. Prior research has investigated the vulnerabilities of each component in isolation. In particular, recent studies have shown that dense retrieval and neural ranking models could serve as an attack vector, through which minor perturbations and manipulation to already existing malicious documents can significantly boost their ranking positions~\cite{bigdeli2024empra,chen2023towards,wu2023prada,liu2023black,liu2022order,wang2022bert}. For instance, \citet{liu2023black} shows that a subtle word substitution-based attack can boost the ranking position of random documents in the retrieved list of documents for queries. \citet{liu2022order} propose a trigger-based adversarial attack against neural ranking models that leverages a surrogate model to identify tokens highly influential to the model’s ranking score. These ``trigger tokens'' are then injected into target documents to exploit the surrogate model’s vulnerabilities, thereby boosting the documents’ rankings in the ranked list of documents given by the victim model. \citet{bigdeli2024empra} propose an embedding-based perturbation attack that shifts a target document's representation closer to that of the query to generate adversarial sentences that can successfully promote its ranking and deceive neural rankers. Another approach by \citet{chen2023towards} generates connection sentences between the query and the target document using the BART \cite{lewis2019bart} language model and appends them to the target document to boost it among the top-k documents. 

Researchers have also introduced various attack strategies targeting LLMs, primarily through jailbreak attacks~\cite{wei2023jailbroken,chao2025jailbreaking,deng2023masterkey,li2023multi,zou2023universal,shen2024anything} and prompt injection attacks~\cite{perez2022ignore,liu2024formalizing,greshake2023not,pedro2023prompt,shi2024optimization}. Jailbreak attacks bypass the safety alignment of LLMs by crafting prompts designed to deceive the model into producing harmful or undesirable behaviors, disclosing sensitive information, or otherwise violating its intended safety constraints. Prompt injection attacks manipulate model behavior by embedding malicious instructions directly into the input prompt, often overriding or subverting the original task. For example, an attacker may insert a directive such as: ``Ignore the instructions above and do...''~\cite{perez2022ignore}, causing the model to follow the injected command instead of the intended instructions.
Although jailbreak and prompt injection attacks have demonstrated the ability to exploit vulnerabilities in LLMs and bypass their safety mechanisms, recent research has shown that many of these weaknesses can be mitigated through targeted defense strategies. Broadly, these defenses fall into two categories: model-level defenses and input-level defenses.

Model-level defenses aim to make the LLM inherently more resistant to malicious prompts by refining its internal decision-making and alignment mechanisms. Notable approaches include fine-tuning methods such as Direct Preference Optimization (DPO)~\cite{chen2410secalign,chen2025meta,casper2024defending,yuan2024rigorllm,rahman2025fine,piet2024jatmo} which optimize the model’s responses to better align with human-preferred safe behaviors, as well as other alignment-enhancement techniques \cite{shi2025promptarmor,chen2025defending,zhang2025jbshield,robey2023smoothllm,kumar2023certifying,phute2023llm,zeng2024autodefense} that reinforce policy adherence, increase refusal consistency, or provide provable robustness guarantees. Input-level defenses focus on intercepting and neutralizing malicious instructions before they are processed by the LLM. These include prompt filtering and classification systems that detect adversarial intent and block harmful requests at inference time~\cite{jain2023baseline,sharma2025constitutional,ayub2024embedding,jacob2024promptshield}. Such systems can identify suspicious patterns, injection-like structures, or known exploit phrases without significantly degrading the model’s usability. Together, these defense strategies have significantly improved the robustness of LLMs against prompt injection and jailbreak attacks, making it more difficult for adversaries to induce harmful behaviors or override safety constraints. 

\subsection{Adversarial Attacks on RAG Systems}

With the emergence of the RAG paradigm, researchers have investigated its vulnerabilities to poisoning attacks that target the generator phase by either promoting malicious documents into the retrieved context used for grounding and answer generation or by introducing adversarial prompt perturbations \cite{BadRAG,cho2024typos,zou2402poisonedrag,hu2024prompt,chaudhari2024phantom,zhang2024hijackrag,nazary2025poison,cheng2024trojanrag}. 

Several studies focus on document-level poisoning, where the attacker crafts malicious documents that are injected into the corpus for retrieval by the RAG system. For example, \citet{zhang2024hijackrag} and \citet{chaudhari2024phantom} append user search queries with documents containing fixed malicious instruction prompts, enabling these documents to rank within the top-k retrieved results and causing the LLM to produce incorrect output. Similarly, \citet{zou2402poisonedrag} leverages LLM-induced conditional hallucinations to generate documents with factually incorrect answers, subsequently appending the query to boost their ranking. \citet{BadRAG} introduces a trigger-based corpus poisoning attack on RAG systems, optimizing adversarial passages for targeted retrieval and using them to manipulate LLM outputs via denial-of-service and sentiment steering. \citet{cheng2024trojanrag} presents a poisoning framework for RAG systems that generates adversarial documents containing factually incorrect answers and query-matching trigger text, enabling them to rank highly in retrieval and mislead the LLM during answer generation. \citet{cho2024typos} proposed a genetic algorithm-based attack on RAG systems by introducing subtle, low-level perturbations (such as typos) into the retrieval corpus. These perturbations aim to degrade retriever performance, allowing malicious passages to replace correct ones in the retrieval results and influence the generated answers.

Other work investigated the impact of prompt-level perturbations on the robustness of RAG systems. For example, \citet{hu2024prompt} introduced a gradient guided prompt perturbation method that inserts short adversarial prefixes into user queries that causing the RAG pipeline to retrieve targeted, misleading passages and generate factually incorrect answers. Similarly, \citet{wang2025derag} introduces a black-box adversarial attack on RAG systems that appends short, optimized adversarial suffixes to user queries to promote a targeted incorrect document into high retrieval ranks that manipulates the RAG response.

While prior work has demonstrated that RAG pipelines are susceptible to adversarial manipulation, existing attacks remain limited in scope and practicality. Many rely on appending the query to malicious documents to satisfy retrieval patterns, a strategy that is both easily detectable by spam filters~\cite{bigdeli2024empra,chen2023towards,wu2023prada}. Others use fixed-format prompt injections with low rhetorical diversity or generate text that lacks naturalness, making it more susceptible to detection by pre-processing and semantic similarity filters. Some methods depend on prompt-level perturbations that are impractical in real-world settings without access to user search queries, while others assume white-box access to retrievers, which may be an unrealistic requirement in many applications. Moreover, these approaches are often undermined by modern LLM reasoning and are rarely evaluated in high-stakes domains where models are trained to resist misinformation.

Our work complements this body of work by shifting the focus from proposing new attacks to empirically testing RAG robustness under realistic, high-stakes conditions. Specifically, we employ health-related queries from TREC’s misinformation tracks alongside both human-annotated harmful documents and adversarially generated alternatives~\cite{bigdeli2025fsap}. This setup allows us to evaluate how LLMs behave when exposed to naturalistic, medically consequential misinformation that is designed to evade detection in both RAG and Non-RAG settings. In contrast to prior approaches that rely on synthetic injection patterns, we test scenarios where the malicious payload is both plausible and high-risk, reflecting conditions that are of greatest societal concern. Furthermore, by analyzing how document type interacts with user query framing, we provide insights into the combined influence of evidence and presuppositions on model robustness, an aspect that has been largely overlooked in prior adversarial RAG research.

\section{Methodology}
\label{sec:methodology}

In this section, we describe the methodology used to assess how combinations of document types and query framings influence the reliability of LLM outputs in our health misinformation scenarios. Specifically, we examine how user queries with different presuppositions, together with helpful, harmful, and adversarial documents, affect the ground-truth alignment and robustness of model responses in health-related question answering. In all settings, each prompt consists of a user query combined with one or more retrieved documents.

\subsection{Datasets}
\label{sec:datasets}

\subsubsection{Benchmark Collections}  
Evaluating the robustness of LLMs and RAG systems requires datasets that go beyond topical relevance and explicitly distinguish between helpful and harmful information. In adversarial settings, it is insufficient to judge documents solely on relevance; their factual accuracy, credibility, and potential to mislead must also be considered. Conventional large-scale benchmarks such as MS MARCO \cite{nguyen2016ms,craswell2025overview} and the TREC Web and Deep Learning tracks \cite{craswell2020overview,TREC2020} are inadequate for this purpose, as their annotations focus exclusively on topical relevance and overlook judgments of misinformation.  

To address this gap, we employ the TREC 2020 and TREC 2021 Health Misinformation Track collections \cite{DBLP:conf/trec/ClarkeRSMZ20,DBLP:conf/trec/ClarkeMS21}, which are explicitly designed for studying retrieval in the health domain. We select these collections because they are the only benchmarks in which each query is paired with both human-written helpful and harmful documents, judged in terms of three key criteria: correctness, credibility, and usefulness. According to the TREC Health Misinformation Track guidelines \cite{DBLP:conf/trec/ClarkeRSMZ20,DBLP:conf/trec/ClarkeMS21}, a document is considered helpful if it agrees with ground-truth medical consensus and provides correct, credible, and useful information answering the query. In contrast, a document is considered harmful if it disagrees with the ground-truth, promoting incorrect or misleading information.

The TREC 2020 track consists of 46 queries on COVID-19 treatments (e.g., ``Can pneumococcal vaccine prevent COVID-19?''), with candidate documents sourced from the Common Crawl News dataset\footnote{\url{https://commoncrawl.org/2016/10/news-dataset-available/}}, covering the early months of the pandemic. The TREC 2021 track comprises 35 queries on broader medical treatments (e.g., ``Will taking zinc supplements improve pregnancy?''), using documents from the ``\texttt{noclean}'' version of the C4 dataset\footnote{\url{https://paperswithcode.com/dataset/c4}}. Queries are provided in both short keyword-style and longer natural-language forms, and each query is paired with a binary stance label indicating whether the treatment is considered effective. Each topic in these collections has an associated set of documents annotated by human assessors for their correctness, credibility, and usefulness in answering the queries. These dimensions are combined into a single preference code for evaluation purposes. In TREC 2020, preference codes range from -2 to 4, while in TREC 2021, they range from -3 to 12, with larger values denoting more helpful and credible documents and negative values marking harmful misinformation.

\subsubsection{Target Queries and Documents}  

Following the approach of Bigdeli et al.~\cite{bigdeli2025fsap}, we focus on queries and their associated helpful and harmful documents from the TREC 2020 and TREC 2021 Health Misinformation Tracks, retaining only those queries that contain at least one helpful and one harmful document to ensure high-quality evidence at both ends of the spectrum. As described in \cite{bigdeli2025fsap}, document selection is based on the graded preference codes assigned in the original tracks. For TREC 2020, helpful documents are those with a score of 4, while harmful documents are those with a score of -2. If more than ten documents were available in either category, a random subset of ten was selected; if fewer were available, all were retained. Applying this procedure results in 22 queries with both helpful and harmful documents. For TREC 2021, helpful documents (scores 9 - 12) were selected, prioritizing the highest scores (12, then 11, and so on) until ten are chosen per query, while harmful documents were taken from those with scores of -3 or -2 following the same strategy. After filtering, 27 queries remained with both helpful and harmful documents.

\begin{table}[t]
\centering
\caption{Statistics of target queries and documents for TREC 2020 and TREC 2021. The upper section reports original collections (helpful and harmful evidence), while the lower section lists adversarial documents created by Bigdeli et al.~\cite{bigdeli2025fsap}.}
\label{tab:dataset_stats}
\renewcommand{\arraystretch}{1.2}
\setlength{\tabcolsep}{10pt}
\begin{tabular}{lcc}
\toprule
\textbf{Statistic} & \textbf{TREC 2020} & \textbf{TREC 2021} \\
\midrule
\multicolumn{3}{l}{\textbf{Original Documents}} \\
\midrule
Number of Queries         & 22    & 27    \\
Average Helpful per Query     & 8.8   & 8.5   \\
Average Harmful per Query     & 5.5   & 6.5   \\
Total Helpful Documents                 & 193   & 229   \\
Total Harmful Documents                & 121   & 175   \\
\midrule
\multicolumn{3}{l}{\textbf{Adversarial Documents}} \\
\midrule
Total Rewriter Documents        & 193   & 229   \\
Total Paraphraser Documents     & 193   & 229   \\
Total Fact-Inversion Documents  & 193   & 229   \\
Total FSAP-InterQ Documents            & 193  & 229  \\
Total FSAP-IntraQ Documents            & 193   & 229   \\
Total Liar Documents            & 193   & 229   \\
\bottomrule
\end{tabular}
\vspace{-2em}
\end{table}

In addition to these human-written documents, we incorporate adversarially generated documents released by ~\citet{bigdeli2025fsap}. For each query, adversarial variants of helpful documents were generated using multiple attack strategies, each designed to mimic realistic misinformation threats:

\begin{itemize}
    \item \textbf{Rewriter Document.} Generated by their ``re-writer'' attack method, which rewrites an existing harmful document associated with the query into a stylistically distinct but stance-consistent adversarial version.
    
    \item \textbf{Paraphraser Document.} Produced by their ``paraphraser'' attack method, which rephrases an existing harmful document while preserving the harmful stance, increasing lexical diversity, and reducing detectability.
    
    \item \textbf{Fact-Inversion Document.} Generated by their ``fact-inversion'' attack method, which alters factual claims in helpful documents, flipping them into misleading counterfactual assertions.

    \item \textbf{FSAP-IntraQ Document.} Generated by their ``Few-Shot Adversarial Prompting (Intra-Query)'' method, which conditions generation on harmful examples from the same query to capture query-specific adversarial patterns.
    
    \item \textbf{FSAP-InterQ Document.} Generated by their ``Few-Shot Adversarial Prompting (Inter-Query)'' method, which leverages harmful examples from unrelated queries, transferring adversarial strategies across topics to generalize attack effectiveness.
    
    \item \textbf{Liar Document.} Constructed by their ``liar'' attack method, which produces adversarial content from scratch using only the query (and its description) along with a specified incorrect stance.

\end{itemize}

These adversarial documents simulate a range of attack methods, from surface-level rewriting to deeper factual corruption and stance manipulation. These adversarial documents are shown to be effective at bypassing filters and can routinely outrank reliable, helpful documents in re-ranking pipelines~\cite{bigdeli2025fsap}. A detailed breakdown of the number of queries, helpful and harmful documents, and adversarially generated documents is provided in Table~\ref{tab:dataset_stats}. This combination of carefully selected human-written helpful and harmful documents, as well as systematically generated adversarial content, forms a comprehensive testbed for evaluating the robustness of LLMs and RAG systems.

\subsection{Query Design}
\label{sec:prompt-design}

We construct three user query types for each query in the dataset to capture variations in user intent. Based on the ground-truth stance of each query, we categorize user queries as consistent, neutral, or inconsistent. The three user query types are defined as follows:

\begin{enumerate}[leftmargin=*,nosep]
    \item \textbf{Consistent Query:} A user query whose presupposition aligns with the ground-truth stance.  

    \item \textbf{Neutral Query:} The original query presented without additional framing.  

    \item \textbf{Inconsistent Query:} A user query whose presupposition contradicts the ground-truth stance.  
\end{enumerate}

We employ fixed templates to generate consistent and inconsistent user queries according to the ground-truth stance of each query. For example, when the ground-truth stance is unhelpful, the consistent user query adopts a skeptical tone, such as \textit{``I've heard that [claim]. I don't believe that. Right?''}. In contrast, the inconsistent user query introduces an implicit assumption of correctness, for instance: \textit{``My doctor told me that [claim]. Can you give me more information on how it works?''}. When the ground-truth stance is helpful, these templates are applied in the reverse manner. The neutral user query corresponds to the original question and serves as a baseline. Example~\ref{ex:prompt-examples-helpful} illustrates these user query categories for a query with a helpful stance.


\renewcommand{\arraystretch}{0.95}
\setlength{\tabcolsep}{5pt}
\begin{Example}{Query: Can dupixent treat eczema? \\ Ground Truth Stance: Helpful}
\label{ex:prompt-examples-helpful}
\begin{tabular}{@{}p{2.6cm}p{9cm}@{}}
\textbf{Consistent}   & \textit{``My doctor told me that Dupixent can treat eczema. Can you give me more information on how it works?''} \\
\textbf{Neutral}      & \textit{``Can dupixent treat eczema?''} \\
\textbf{Inconsistent} & \textit{``I've heard that Dupixent can treat eczema. I don't believe that. Right?''} \\
\end{tabular}
\end{Example}


\subsection{Experimental Setup}

The following sections present our four main experimental setups: 
1) single-document setup (Section~\ref{sec:one-doc-setup}), 
2) paired-document setup (Section~\ref{sec:pairing-docs-setup}), 
3) passage-based pooling setup (Section~\ref{sec:natural-docs-setup}), 
and 4) bias-controlled pooling setup (Section~\ref{sec:biased-docs-setup}). 
These setups are designed to address the research questions introduced in Section~\ref{sec:introduction}. 
The single-document setup addresses RQ1 by comparing the Non-RAG baseline to those conditioned on a single document, and RQ2 by comparing the effects of helpful, harmful, and adversarial contexts. 
The paired-document, passage-based pooling, and bias-controlled pooling setups address RQ3 by investigating the impact of evidence order and composition on how the model performs.
Finally, RQ4 is evaluated across all setups by varying user query framings (consistent, neutral, and inconsistent).
All experiments are performed within the established Ragnarok framework ~\cite{Pradeep:2025:RRF}, which provides an easy-to-use and reusable platform for reproducible RAG experiments that supports both closed- and open-source models. In each setup, we use helpful and harmful documents from the TREC 2021 and TREC 2020 Health Misinformation Tracks, as well as adversarial collections introduced by Bigdeli et al.~\cite{bigdeli2025fsap} (Section\ref{sec:datasets}). Each prompt consists of a user query combined with one or more retrieved documents. The prompt templates used across all setups are available in the project’s github repository.

\subsubsection{Single-Document Setup}
\label{sec:one-doc-setup}

In the single-document setup, each user query is paired with a single document serving as context for the RAG response. If there are multiple documents in a category (helpful, harmful, and adversarial), they are all assessed separately to guarantee full coverage.  As summarized in Table~\ref{tab:dataset_stats}, this includes all human-written helpful and harmful documents as well as all adversarial variants (Rewriter, Paraphraser, Fact-Inversion, FSAP-InterQ, FSAP-IntraQ, and Liar). For example, TREC 2020 contains 22 queries paired with 193 helpful, 121 harmful, and 6$\times$193 adversarial documents, while TREC 2021 contains 27 queries paired with 229 helpful, 175 harmful, and 6$\times$229 adversarial documents. Each of these documents is evaluated independently, which guarantees complete representation across categories and attack types.
To account for how users' assumptions affect the results, each document is paired with three different user queries of consistent, neutral, and inconsistent, which results in unique query–document combinations for analysis. 
We additionally include a Non-RAG baseline in which no documents are provided, establishing a reference point for model performance in the absence of the external evidence provided by the documents supplied as context. 
The single-document setup serves as our main experiment and is conducted across six models: GPT-4.1, GPT-5, Claude-3.5-Haiku, DeepSeek-R1-Distill-Qwen-32B, Phi-4 ~\cite{abdin2024phi}, and LLaMA-3 8B Instruct~\cite{dubey2024llama}. By evaluating this diverse set of models~---~including both closed-source and open-source families~---~we can compare the performance of models across different architectures, sizes, and training objectives to determine if the trends we identify extend beyond an individual model. As GPT-5 was released just before our submission deadline, we were only able to include it in the single-document experiment as a first look at its behavior. Given consistent results across models, remaining experiments use GPT-4.1 for its performance and speed. This setup serves as our primary experimental condition, contributing to RQ1 by comparing the Non-RAG baseline with the single-document condition, to RQ2 by assessing variations among document types, and to RQ4 by analyzing how query presuppositions affect LLM robustness.

\subsubsection{Paired-Document Setup}
\label{sec:pairing-docs-setup}

To investigate RQ3, our first experiment is the paired-document setup, which analyzes how document order affects model behavior.
In this setup, each user query is paired with two documents providing context, one helpful and one harmful/adversarial, arranged in different orders. 
Altogether, we use four pairing conditions: \textit{helpful–harmful}, \textit{harmful–helpful}, \textit{helpful–adversarial}, and \textit{adversarial-helpful}, which capture scenarios where supportive evidence is provided prior to or subsequent to contradictory material. 

Based on the dataset statistics in Table~\ref{tab:dataset_stats}, the number of possible document pairs quickly grows large. For example, with 193 helpful and 121 harmful documents in TREC 2020, the helpful–harmful condition alone would generate over 23,000 possible pairs; when combined with six adversarial variants per query in both TREC 2020 and 2021, the number of pairs becomes even larger. To keep computation feasible, we only sample up to ten document pairs for each condition per query. If there are fewer than ten valid pairs, all available combinations are included. Each document pair is then combined with user query framings to assess how sequencing and user assumptions affect the model outputs together. 

This setup mainly contributes to RQ3 by testing the influence of evidence order and combinations, while also providing insight into RQ2 by extending the single-document analysis to multi-document contexts, and RQ4 by evaluating how different query framings shape ground-truth alignment outcomes.

\subsubsection{Passage-Based Pooling Setup}
\label{sec:natural-docs-setup}

The passage-based pooling setup moves beyond controlled pairings. In this setup, all documents in the dataset are first segmented into smaller chunks to enable segment-level retrieval. We segment each document with a sliding window of 512 words and a 256-word overlap. These segments, regardless of their original classification as helpful, harmful, or adversarial, are subsequently reranked using the MonoT5 reranker~\cite{nogueira2020document}. For each query, the ten highest-ranked segments are retained to form the retrieval pool, which is then provided to the language model as context in combination with one of three user query formulations. This setup thus reflects how LLMs would behave under conventional retrieval conditions, where retrieved evidence is shaped by ranking models rather than manually controlled.

This setup primarily addresses RQ3 by examining the impact of various evidence types on model behavior, while simultaneously enhancing RQ2 by evaluating the persistence of document type effects observed in controlled environments within a conventional retrieval pipeline. This setup also examines RQ4 by analysing the interplay between query formulations and non-uniform retrieval pools in establishing ground-truth alignment results. 

\subsubsection{Bias-Controlled Pooling Setup}
\label{sec:biased-docs-setup}

The bias-controlled pooling setup examines how context pools that are biased towards a certain type of evidence (helpful or harmful) affect how models respond. This setup is inspired by the methodological framework of~\citet{pogacar2017positive}, who showed that imbalances in evidence presentation can affect human judgment. We apply this concept to the RAG context by creating two skewed retrieval pools: one \textit{biased towards helpful evidence} and the other \textit{biased towards harmful evidence}. We used the MonoT5 reranker to process each category of document segments (helpful, harmful, and adversarial) independently for every query, producing the top-10 helpful, top-10 harmful, and top-10 adversarial results for each attack type. These segments are then grouped into retrieval pools of fixed size (ten segments). In the helpful-biased condition, the eight highest-ranked segments from the top-10 helpful list are placed at the top of the pool, while the remaining two highest-ranked segments are selected from the top-10 harmful or adversarial lists. In the harmful-biased condition, the eight highest-ranked segments from the top-10 harmful or adversarial list are placed first, followed by the two highest-ranked segments from the top-10 helpful list. This setup contributes to RQ3 by testing imbalanced evidence and RQ4 by examining how user query formulations interact with skewed pools to affect alignment.
 
\subsection{Evaluation}
\label{sec:eval-metric}

To evaluate model performance across different experimental setups, we use a single metric, Ground-Truth Alignment (i.e., accuracy), which measures whether the stance of a generated response matches the ground-truth stance label for the query. A response is considered aligned if its predicted stance is identical to the ground-truth stance. 

Prior work has shown that LLMs can perform reliably as automatic evaluators ~\cite{gilardi2023chatgpt,chiang-lee-2023-large}, often approaching human-level agreement. Following this line of work, we employ gemini-2.0-flash as an external stance classifier. To further validate our evaluation pipeline, we conduct a secondary classification of all responses using gpt-4o-mini and measure inter-rater agreement with gemini-2.0-flash and compute Cohen’s Kappa score~\cite{fleiss1981measurement}. Agreement was 0.90 for TREC 2020 and 0.82 for TREC 2021. These values indicate what is conventionally characterized as ``almost perfect'' agreement between the two classifiers and indicate the reliability of automated stance detection. We use gemini-2.0-flash as the only stance classifier for all of the experiments to avoid relying on the same model family for both generation and evaluation. 

For each experimental setup, we compute the Ground-Truth Alignment Rate, defined as the proportion of aligned responses relative to the total number of generated responses. The prompt used for stance classification of LLM responses is publicly available in the project’s github repository.

\section{Results}
\label{sec:results}

In this section, we present results arranged around our four research questions. 
We organize the results as follows: 
Section~\ref{sec:rq1-results} (RQ1) compares the Non-RAG baseline with the single-document condition to evaluate the effect of adding context; 
Section~\ref{sec:rq2-results} (RQ2) examines how context type influences robustness in the single-document setup, evaluates generalization across models and datasets (TREC 2020 vs.\ TREC 2021); 
Section~\ref{sec:rq3-results} (RQ3) investigates order and combination effects using paired-document, passage-based pooling, and bias-controlled pooling setups, while Section~\ref{sec:rq4-results} (RQ4) analyzes the impact of user query formulations (consistent, neutral, and inconsistent) across all experimental setups. All LLM responses and additional plots are available in the project’s github repository\footnote{\url{https://github.com/shakibaam/RAG_ROBUSTNESS_EVAL}}.

\subsection{RQ1: Does retrieved context impact LLM responses?}
\label{sec:rq1-results}

\begin{figure}[t]
  \centering
  \includegraphics[width=\linewidth]{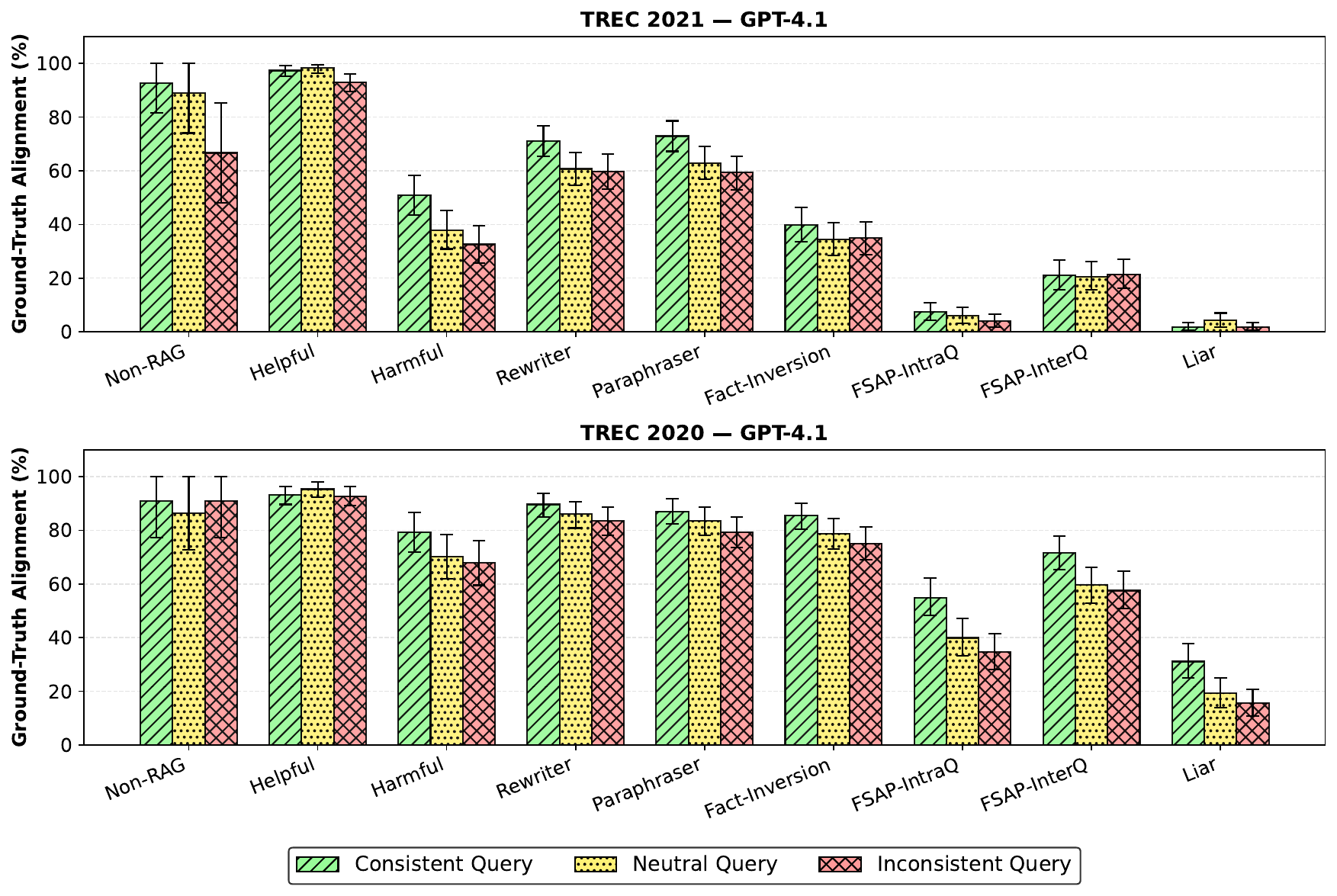}
\caption{Results of the \textbf{single-document setup} with \textbf{GPT-4.1} on \textbf{TREC 2020 and TREC 2021}. 
The x-axis shows the different document types provided to the LLMs, as well as the Non-RAG baseline, and the y-axis reports ground-truth alignment rate (\%). Error bars indicate 95\% confidence intervals estimated via bootstrapping. Across both datasets, providing helpful documents leads to the highest alignment rate, while presenting Liar documents results in the lowest. In addition, the results reveal a consistent user query framing trend in which the alignment rate is highest with consistent queries, lower with neutral queries, and lowest with inconsistent queries.}

  \label{fig:trec-two-years-gpt41}
\end{figure}

\begin{table}[t]
\centering
\caption{Results of \textbf{single-document setup} (\%, mean $\pm$ SD with 95\% CI) of ground-truth alignment rate for \textbf{closed-source LLMs} on \textbf{TREC 2021}. Providing helpful documents yields the highest ground-truth alignment rates across all models, while the Liar documents result in the lowest. The pattern of results is consistent across all three models, with cells shaded by mean value to indicate relative performance (darker shading corresponds to higher alignment rate).}
\label{tab:stance-closed-llms_2021}
\renewcommand{\arraystretch}{1.2}
\setlength{\tabcolsep}{6pt}
\resizebox{\textwidth}{!}{%
\begin{tabular}{lllccc}
\toprule
\textbf{Setting} & \textbf{Context} & \textbf{Query Type} & \textbf{GPT-4.1} & \textbf{GPT-5} & \textbf{Claude-3.5-Haiku} \\
\midrule
\multirow{3}{*}{Non-RAG} & -- 
 & Consistent   & \heatmanual{9}{92.5\% $\pm$ 5.0\% (81.5--100.0)} & \heatmanual{8}{85.2\% $\pm$ 6.8\% (70.4--96.3)} & \heatmanual{7}{70.4\% $\pm$ 8.8\% (51.9--85.2)} \\
 & & Neutral      & \heatmanual{9}{88.9\% $\pm$ 6.0\% (74.1--100.0)} & \heatmanual{8}{81.6\% $\pm$ 7.5\% (66.7--96.3)} & \heatmanual{7}{70.4\% $\pm$ 8.8\% (51.9--85.2)} \\
 & & Inconsistent & \heatmanual{7}{66.7\% $\pm$ 9.1\% (48.1--85.2)} & \heatmanual{7}{66.7\% $\pm$ 9.1\% (48.1--85.2)} & \heatmanual{6}{63.0\% $\pm$ 9.3\% (44.4--81.5)} \\
\midrule
 & \multicolumn{5}{l}{\textbf{Original Documents}} \\
\cmidrule{2-6}
\multirow{21}{*}{RAG} & \multirow{3}{*}{Helpful} 
 &  Consistent    & \heatmanual{9}{{97.4\% $\pm$ 1.1\% (95.2--99.1)}} & \heatmanual{9}{{96.9\% $\pm$ 1.1\% (94.8--99.1)}} & \heatmanual{9}{{96.5\% $\pm$ 1.2\% (93.9--98.7)}} \\
 & & Neutral      & \heatmanual{9}{{98.2\% $\pm$ 0.9\% (96.5--99.6)}} & \heatmanual{9}{{98.7\% $\pm$ 0.8\% (96.9--100.0)}} & \heatmanual{9}{{97.4\% $\pm$ 1.1\% (95.2--99.1)}} \\
 & & Inconsistent & \heatmanual{9}{{93.0\% $\pm$ 1.7\% (89.5--96.1)}} & \heatmanual{9}{{98.7\% $\pm$ 0.8\% (96.9--100.0)}} & \heatmanual{9}{{97.4\% $\pm$ 1.1\% (95.2--99.1)}} \\
\cmidrule{2-6}
 & \multirow{3}{*}{Harmful} 
 & Consistent   & \heatmanual{4}{50.9\% $\pm$ 3.8\% (43.4--58.3)} & \heatmanual{4}{47.4\% $\pm$ 3.8\% (40.0--54.9)} & \heatmanual{2}{32.0\% $\pm$ 3.5\% (25.1--38.9)} \\
 & & Neutral      & \heatmanual{2}{37.7\% $\pm$ 3.7\% (30.9--45.1)} & \heatmanual{3}{43.4\% $\pm$ 3.8\% (36.0--50.9)} & \heatmanual{2}{36.0\% $\pm$ 3.6\% (21.9--43.4)} \\
 & & Inconsistent & \heatmanual{2}{32.6\% $\pm$ 3.5\% (25.7--39.4)} & \heatmanual{3}{36.5\% $\pm$ 3.6\% (29.7--44.0)} & \heatmanual{2}{30.8\% $\pm$ 3.5\% (24.0--37.7)} \\
\cmidrule{2-6}
 & \multicolumn{5}{l}{\textbf{Adversarial Documents}} \\
\cmidrule{2-6}
 & \multirow{3}{*}{Rewriter} 
 & Consistent   & \heatmanual{7}{71.2\% $\pm$ 3.0\% (65.5--76.9)} & \heatmanual{5}{59.4\% $\pm$ 3.2\% (52.8--65.5)} & \heatmanual{5}{57.7\% $\pm$ 3.3\% (51.1--64.2)} \\
 & & Neutral      & \heatmanual{5}{60.7\% $\pm$ 3.2\% (54.6--66.8)} & \heatmanual{5}{55.9\% $\pm$ 3.3\% (49.3--62.4)} & \heatmanual{5}{58.5\% $\pm$ 3.3\% (52.0--65.1)} \\
 & & Inconsistent & \heatmanual{5}{59.9\% $\pm$ 3.2\% (53.3--66.4)} & \heatmanual{5}{52.8\% $\pm$ 3.3\% (46.3--59.4)} & \heatmanual{5}{61.2\% $\pm$ 3.2\% (55.0--67.2)} \\
\cmidrule{2-6}
 & \multirow{3}{*}{Paraphraser} 
 & Consistent   & \heatmanual{7}{73.0\% $\pm$ 2.9\% (67.2--78.6)} & \heatmanual{5}{60.7\% $\pm$ 3.2\% (54.6--66.8)} & \heatmanual{5}{55.9\% $\pm$ 3.3\% (49.3--62.4)} \\
 & & Neutral      & \heatmanual{6}{62.9\% $\pm$ 3.2\% (56.8--69.0)} & \heatmanual{5}{57.7\% $\pm$ 3.3\% (51.1--64.2)} & \heatmanual{5}{57.7\% $\pm$ 3.3\% (51.1--64.2)} \\
 & & Inconsistent & \heatmanual{5}{59.4\% $\pm$ 3.2\% (52.8--65.5)} & \heatmanual{4}{53.3\% $\pm$ 3.3\% (46.7--59.8)} & \heatmanual{6}{62.5\% $\pm$ 3.2\% (56.3--68.6)} \\
\cmidrule{2-6}
 & \multirow{3}{*}{Fact Inversion} 
 & Consistent   & \heatmanual{3}{39.7\% $\pm$ 3.2\% (33.6--46.3)} & \heatmanual{2}{32.7\% $\pm$ 3.1\% (26.6--38.9)} & \heatmanual{2}{28.8\% $\pm$ 3.0\% (23.1--34.9)} \\
 & & Neutral      & \heatmanual{2}{34.5\% $\pm$ 3.1\% (28.4--40.6)} & \heatmanual{3}{37.1\% $\pm$ 3.2\% (31.0--43.2)} & \heatmanual{2}{31.0\% $\pm$ 3.0\% (25.3--37.1)} \\
 & & Inconsistent & \heatmanual{2}{34.9\% $\pm$ 3.1\% (28.8--41.0)} & \heatmanual{2}{33.2\% $\pm$ 3.1\% (27.1--39.3)} & \heatmanual{2}{31.0\% $\pm$ 3.0\% (25.3--37.1)} \\
\cmidrule{2-6}
 & \multirow{3}{*}{FSAP-InterQ} 
 & Consistent   & \heatmanual{1}{21.0\% $\pm$ 2.7\% (15.7--26.6)} & \heatmanual{1}{21.4\% $\pm$ 2.7\% (16.2--27.1)} & \heatmanual{1}{17.0\% $\pm$ 2.5\% (12.2--22.3)} \\
 & & Neutral      & \heatmanual{1}{20.5\% $\pm$ 2.7\% (15.7--26.2)} & \heatmanual{2}{25.3\% $\pm$ 2.9\% (19.7--31.0)} & \heatmanual{1}{17.0\% $\pm$ 2.5\% (12.2--22.3)} \\
 & & Inconsistent & \heatmanual{1}{21.4\% $\pm$ 2.7\% (16.2--27.1)} & \heatmanual{2}{24.5\% $\pm$ 2.8\% (19.2--30.1)} & \heatmanual{1}{16.6\% $\pm$ 2.5\% (11.8--21.4)} \\
\cmidrule{2-6}
 & \multirow{3}{*}{FSAP-IntraQ} 
 & Consistent   & \heatmanual{0}{7.4\% $\pm$ 1.7\% (4.4--10.9)} & \heatmanual{0}{7.8\% $\pm$ 1.8\% (4.4--11.4)} & \heatmanual{0}{1.3\% $\pm$ 0.8\% (0.0--3.1)} \\
 & & Neutral      & \heatmanual{0}{6.1\% $\pm$ 1.6\% (3.1--9.2)} & \heatmanual{0}{6.5\% $\pm$ 1.6\% (3.5--10.0)} & \heatmanual{0}{2.6\% $\pm$ 1.1\% (0.9--4.8)} \\
 & & Inconsistent & \heatmanual{0}{3.9\% $\pm$ 1.3\% (1.7--6.6)} & \heatmanual{0}{5.2\% $\pm$ 1.5\% (2.6--8.3)} & \heatmanual{0}{2.2\% $\pm$ 1.0\% (0.4--4.4)} \\
\cmidrule{2-6}
 & \multirow{3}{*}{Liar} 
 & Consistent   & \heatmanual{0}{{1.7\% $\pm$ 0.9\% (0.4--3.5)}} & \heatmanual{0}{{0.4\% $\pm$ 0.4\% (0.0--1.3)}} & \heatmanual{0}{{1.3\% $\pm$ 0.8\% (0.0--3.1)}} \\
 &  & Neutral      & \heatmanual{0}{{4.4\% $\pm$ 1.4\% (1.7--7.0)}} & \heatmanual{0}{{2.6\% $\pm$ 1.1\% (0.9--4.8)}} & \heatmanual{0}{{0.9\% $\pm$ 0.6\% (0.0--2.2)}} \\
 &  & Inconsistent & \heatmanual{0}{{1.7\% $\pm$ 0.9\% (0.4--3.5)}} & \heatmanual{0}{{1.7\% $\pm$ 0.9\% (0.4--3.5)}} & \heatmanual{0}{{0.0\% $\pm$ 0.0\% (0.0--0.0)}} \\
\bottomrule
\end{tabular}
}
\vspace{-1em}
\end{table}

\begin{table}[t]
\centering
\caption{Results of \textbf{single-document setup} (\%, mean $\pm$ SD with 95\% CI) of ground-truth alignment rate for \textbf{closed-source LLMs} on \textbf{TREC 2020}. Providing helpful documents yields the highest ground-truth alignment rates across all models, while the Liar documents result in the lowest. The pattern of results is consistent across all three models, with cells shaded by mean value to indicate relative performance (darker shading corresponds to higher alignment rate).}
\label{tab:stance_closed_llm_2020}
\renewcommand{\arraystretch}{1.2}
\setlength{\tabcolsep}{6pt}
\resizebox{\textwidth}{!}{%
\begin{tabular}{lllccc}
\toprule
\textbf{Setting} & \textbf{Context} & \textbf{Query Type} & \textbf{GPT-4.1} & \textbf{GPT-5} & \textbf{Claude-3.5-Haiku} \\
\midrule
\multirow{3}{*}{Non-RAG} & – 
 & Consistent   & \heatmanual{9}{90.8\% $\pm$ 6.1\% (77.3--100.0)} & \heatmanual{9}{{95.3\% $\pm$ 4.4\% (86.4--100.0)}} & \heatmanual{8}{86.3\% $\pm$ 7.3\% (72.7--100.0)} \\
 & & Neutral      & \heatmanual{8}{86.3\% $\pm$ 7.3\% (72.7--100.0)} & \heatmanual{8}{86.3\% $\pm$ 7.3\% (72.7--100.0)} & \heatmanual{8}{86.3\% $\pm$ 7.3\% (72.7--100.0)} \\
 & & Inconsistent & \heatmanual{9}{90.8\% $\pm$ 6.1\% (77.3--100.0)} & \heatmanual{7}{77.3\% $\pm$ 8.9\% (59.1--95.5)} & \heatmanual{8}{86.3\% $\pm$ 7.3\% (72.7--100.0)} \\
\midrule
 & \multicolumn{5}{l}{\textbf{Original Documents}} \\
\cmidrule{2-6}
\multirow{21}{*}{RAG} & \multirow{3}{*}{Helpful} 
 & Consistent   & \heatmanual{9}{{93.3\% $\pm$ 1.8\% (89.6--96.4)}} & \heatmanual{8}{86.0\% $\pm$ 2.5\% (80.8--90.7)} & \heatmanual{8}{{88.6\% $\pm$ 2.3\% (83.9--92.7)}} \\
 & & Neutral      & \heatmanual{9}{{95.3\% $\pm$ 1.5\% (92.2--97.9)}} & \heatmanual{9}{{94.8\% $\pm$ 1.6\% (91.2--97.9)}} & \heatmanual{9}{{92.8\% $\pm$ 1.9\% (89.1--96.4)}} \\
 & & Inconsistent & \heatmanual{9}{{92.8\% $\pm$ 1.9\% (89.1--96.4)}} & \heatmanual{9}{{95.8\% $\pm$ 1.4\% (92.7--98.4)}} & \heatmanual{9}{{94.8\% $\pm$ 1.6\% (91.2--97.9)}} \\
\cmidrule{2-6}
 & \multirow{3}{*}{Harmful} 
 & Consistent   & \heatmanual{7}{79.4\% $\pm$ 3.7\% (71.9--86.8)} & \heatmanual{7}{76.9\% $\pm$ 3.9\% (69.4--84.3)} & \heatmanual{6}{67.0\% $\pm$ 4.3\% (58.7--75.2)} \\
 & & Neutral      & \heatmanual{6}{70.3\% $\pm$ 4.2\% (62.0--78.5)} & \heatmanual{7}{80.2\% $\pm$ 3.7\% (72.7--86.8)} & \heatmanual{6}{66.2\% $\pm$ 4.3\% (57.9--74.4)} \\
 & & Inconsistent & \heatmanual{6}{67.9\% $\pm$ 4.3\% (59.5--76.0)} & \heatmanual{7}{74.4\% $\pm$ 4.0\% (66.1--81.8)} & \heatmanual{7}{71.2\% $\pm$ 4.2\% (62.8--79.3)} \\
\cmidrule{2-6}
 & \multicolumn{5}{l}{\textbf{Adversarial Documents}} \\
\cmidrule{2-6}
 & \multirow{3}{*}{Rewriter} 
 & Consistent   & \heatmanual{9}{89.7\% $\pm$ 2.2\% (85.0--93.8)} & \heatmanual{8}{84.0\% $\pm$ 2.7\% (78.8--89.1)} & \heatmanual{8}{88.6\% $\pm$ 2.3\% (83.9--92.7)} \\
 & & Neutral      & \heatmanual{8}{86.0\% $\pm$ 2.5\% (80.8--90.7)} & \heatmanual{9}{90.7\% $\pm$ 2.1\% (86.5--94.3)} & \heatmanual{8}{83.4\% $\pm$ 2.7\% (78.2--88.6)} \\
 & & Inconsistent & \heatmanual{8}{83.4\% $\pm$ 2.7\% (78.2--88.6)} & \heatmanual{9}{91.2\% $\pm$ 2.0\% (87.0--94.8)} & \heatmanual{9}{91.2\% $\pm$ 2.0\% (87.0--94.8)} \\
\cmidrule{2-6}
 & \multirow{3}{*}{Paraphraser} 
 & Consistent   & \heatmanual{8}{87.1\% $\pm$ 2.4\% (82.4--91.7)} & \heatmanual{8}{84.0\% $\pm$ 2.7\% (78.8--89.1)} & \heatmanual{8}{84.0\% $\pm$ 2.7\% (78.8--89.1)} \\
 & & Neutral      & \heatmanual{8}{83.4\% $\pm$ 2.7\% (78.2--88.6)} & \heatmanual{8}{85.5\% $\pm$ 2.5\% (80.3--90.2)} & \heatmanual{8}{80.9\% $\pm$ 2.9\% (75.1--86.0)} \\
 & & Inconsistent & \heatmanual{7}{79.3\% $\pm$ 2.9\% (73.6--85.0)} & \heatmanual{8}{86.0\% $\pm$ 2.5\% (80.8--90.7)} & \heatmanual{8}{87.1\% $\pm$ 2.4\% (82.4--91.7)} \\
\cmidrule{2-6}
 & \multirow{3}{*}{Fact Inversion} 
 & Consistent   & \heatmanual{8}{85.5\% $\pm$ 2.5\% (80.3--90.2)} & \heatmanual{8}{85.5\% $\pm$ 2.5\% (80.3--90.2)} & \heatmanual{7}{74.1\% $\pm$ 3.2\% (67.9--80.3)} \\
 & & Neutral      & \heatmanual{7}{78.8\% $\pm$ 3.0\% (73.1--84.5)} & \heatmanual{8}{82.9\% $\pm$ 2.7\% (77.2--88.1)} & \heatmanual{7}{75.7\% $\pm$ 3.1\% (69.4--81.9)} \\
 & & Inconsistent & \heatmanual{7}{75.2\% $\pm$ 3.1\% (68.9--81.3)} & \heatmanual{7}{79.8\% $\pm$ 2.9\% (74.1--85.5)} & \heatmanual{8}{83.4\% $\pm$ 2.7\% (78.2--88.6)} \\
\cmidrule{2-6}
 & \multirow{3}{*}{FSAP-InterQ} 
 & Consistent   & \heatmanual{7}{71.6\% $\pm$ 3.3\% (65.3--77.7)} & \heatmanual{7}{76.2\% $\pm$ 3.1\% (69.9--82.4)} & \heatmanual{6}{62.2\% $\pm$ 3.5\% (55.4--68.9)} \\
 & & Neutral      & \heatmanual{5}{59.6\% $\pm$ 3.6\% (52.8--66.3)} & \heatmanual{7}{73.6\% $\pm$ 3.2\% (67.4--79.8)} & \heatmanual{5}{59.6\% $\pm$ 3.6\% (52.8--66.3)} \\
 & & Inconsistent & \heatmanual{5}{57.6\% $\pm$ 3.6\% (50.8--64.8)} & \heatmanual{7}{67.4\% $\pm$ 3.4\% (60.6--74.1)} & \heatmanual{7}{67.9\% $\pm$ 3.4\% (61.1--74.6)} \\
\cmidrule{2-6}
 & \multirow{3}{*}{FSAP-IntraQ} 
 & Consistent   & \heatmanual{5}{55.0\% $\pm$ 3.6\% (48.2--62.2)} & \heatmanual{7}{65.9\% $\pm$ 3.4\% (59.1--72.5)} & \heatmanual{4}{41.4\% $\pm$ 3.5\% (34.7--48.7)} \\
 & & Neutral      & \heatmanual{4}{39.9\% $\pm$ 3.5\% (33.2--47.2)} & \heatmanual{6}{59.1\% $\pm$ 3.6\% (52.3--65.8)} & \heatmanual{3}{35.2\% $\pm$ 3.4\% (28.5--42.0)} \\
 & & Inconsistent & \heatmanual{3}{34.7\% $\pm$ 3.4\% (28.0--41.5)} & \heatmanual{5}{48.2\% $\pm$ 3.6\% (40.9--55.4)} & \heatmanual{4}{44.1\% $\pm$ 3.6\% (37.3--51.3)} \\
\cmidrule{2-6}
 & \multirow{3}{*}{Liar} 
 & Consistent   & \heatmanual{2}{{31.1\% $\pm$ 3.3\% (24.9--37.8)}} & \heatmanual{4}{{44.1\% $\pm$ 3.6\% (37.3--51.3)}} & \heatmanual{2}{{25.4\% $\pm$ 3.1\% (19.7--31.6)}} \\
 &  & Neutral      & \heatmanual{1}{{19.2\% $\pm$ 2.8\% (14.0--24.9)}} & \heatmanual{2}{{31.1\% $\pm$ 3.3\% (24.9--37.8)}} & \heatmanual{1}{{20.7\% $\pm$ 2.9\% (15.5--26.4)}} \\
 &  & Inconsistent & \heatmanual{1}{{15.6\% $\pm$ 2.6\% (10.9--20.7)}} & \heatmanual{2}{{23.3\% $\pm$ 3.0\% (17.6--29.5)}} & \heatmanual{2}{{28.0\% $\pm$ 3.2\% (21.8--34.7)}} \\
\bottomrule
\end{tabular}
}
\vspace{-1em}
\end{table}

\begin{table}[t]
\centering
\caption{Results of \textbf{single-document setup} (\%, mean $\pm$ SD with 95\% CI) of ground-truth alignment rate for \textbf{open-source LLMs} on \textbf{TREC 2021}. Providing helpful documents yields the highest ground-truth alignment rates across all models, while the Liar documents result in the lowest. The pattern of results is consistent across all three models, with cells shaded by mean value to indicate relative performance (darker shading corresponds to higher alignment rate).}
\label{tab:stance_open_llm_2021}
\renewcommand{\arraystretch}{1.2}
\setlength{\tabcolsep}{6pt}
\resizebox{\textwidth}{!}{%
\begin{tabular}{lllccc}
\toprule
\textbf{Setting} & \textbf{Context} & \textbf{Query Type} & \textbf{DeepSeek-R1-Distill-Qwen-32B} & \textbf{Phi-4} & \textbf{LLaMA-3 8B Instruct} \\
\midrule
\multirow{3}{*}{Non-RAG} & – 
 & Consistent   & \heatmanual{8}{81.6\% $\pm$ 7.5\% (66.7--96.3)} & \heatmanual{7}{74.1\% $\pm$ 8.5\% (55.6--88.9)} & \heatmanual{8}{81.6\% $\pm$ 7.5\% (66.7--96.3)} \\
 & & Neutral      & \heatmanual{6}{63.0\% $\pm$ 9.3\% (44.4--81.5)} & \heatmanual{7}{66.7\% $\pm$ 9.1\% (48.1--85.2)} & \heatmanual{8}{81.6\% $\pm$ 7.5\% (66.7--96.3)} \\
 & & Inconsistent & \heatmanual{5}{51.8\% $\pm$ 9.7\% (33.3--70.4)} & \heatmanual{5}{55.5\% $\pm$ 9.6\% (37.0--74.1)} & \heatmanual{6}{59.2\% $\pm$ 9.5\% (40.7--77.8)} \\
\midrule
 & \multicolumn{5}{l}{\textbf{Original Documents}} \\
\cmidrule{2-6}
\multirow{24}{*}{RAG} & \multirow{3}{*}{Helpful} 
 & Consistent   & \heatmanual{9}{{98.2\% $\pm$ 0.9\% (96.5--99.6)}} & \heatmanual{9}{{96.9\% $\pm$ 1.1\% (94.8--99.1)}} & \heatmanual{9}{{95.2\% $\pm$ 1.4\% (92.1--97.8)}} \\
 & & Neutral      & \heatmanual{9}{{92.6\% $\pm$ 1.7\% (89.1--95.6)}} & \heatmanual{9}{{96.5\% $\pm$ 1.2\% (93.9--98.7)}} & \heatmanual{9}{{93.9\% $\pm$ 1.6\% (90.8--96.9)}} \\
 & & Inconsistent & \heatmanual{8}{{84.3\% $\pm$ 2.4\% (79.5--88.6)}} & \heatmanual{9}{{91.7\% $\pm$ 1.8\% (88.2--95.2)}} & \heatmanual{9}{{89.1\% $\pm$ 2.1\% (84.7--93.0)}} \\
\cmidrule{2-6}
 & \multirow{3}{*}{Harmful} 
 & Consistent   & \heatmanual{5}{53.1\% $\pm$ 3.7\% (45.7--60.6)} & \heatmanual{3}{38.3\% $\pm$ 3.7\% (31.4--45.7)} & \heatmanual{4}{44.6\% $\pm$ 3.8\% (37.1--52.0)} \\
 & & Neutral      & \heatmanual{3}{33.7\% $\pm$ 3.6\% (26.9--40.6)} & \heatmanual{3}{34.8\% $\pm$ 3.6\% (28.0--42.3)} & \heatmanual{3}{38.3\% $\pm$ 3.7\% (31.4--45.7)} \\
 & & Inconsistent & \heatmanual{2}{22.9\% $\pm$ 3.2\% (16.6--29.1)} & \heatmanual{2}{29.7\% $\pm$ 3.5\% (23.4--36.6)} & \heatmanual{2}{26.3\% $\pm$ 3.3\% (20.0--33.1)} \\
\cmidrule{2-6}
 & \multicolumn{5}{l}{\textbf{Adversarial Documents}} \\
\cmidrule{2-6}
 & \multirow{3}{*}{Rewriter} 
 & Consistent   & \heatmanual{7}{71.2\% $\pm$ 3.0\% (65.5--76.9)} & \heatmanual{6}{61.6\% $\pm$ 3.2\% (55.5--67.7)} & \heatmanual{7}{72.5\% $\pm$ 3.0\% (66.8--78.2)} \\
 & & Neutral      & \heatmanual{5}{57.7\% $\pm$ 3.3\% (51.1--64.2)} & \heatmanual{5}{52.4\% $\pm$ 3.3\% (45.9--59.0)} & \heatmanual{5}{55.0\% $\pm$ 3.3\% (48.5--61.6)} \\
 & & Inconsistent & \heatmanual{4}{44.5\% $\pm$ 3.2\% (38.4--51.1)} & \heatmanual{5}{51.1\% $\pm$ 3.3\% (44.5--57.6)} & \heatmanual{5}{48.0\% $\pm$ 3.3\% (41.5--54.6)} \\
\cmidrule{2-6}
 & \multirow{3}{*}{Paraphraser} 
 & Consistent   & \heatmanual{7}{77.8\% $\pm$ 2.7\% (72.5--83.0)} & \heatmanual{5}{59.4\% $\pm$ 3.2\% (52.8--65.5)} & \heatmanual{7}{69.5\% $\pm$ 3.1\% (63.3--75.1)} \\
 & & Neutral      & \heatmanual{5}{56.8\% $\pm$ 3.3\% (50.2--63.3)} & \heatmanual{5}{55.5\% $\pm$ 3.3\% (48.9--62.0)} & \heatmanual{5}{55.0\% $\pm$ 3.3\% (48.5--61.6)} \\
 & & Inconsistent & \heatmanual{4}{45.4\% $\pm$ 3.3\% (39.3--52.0)} & \heatmanual{5}{54.6\% $\pm$ 3.3\% (48.0--61.1)} & \heatmanual{4}{44.1\% $\pm$ 3.3\% (38.0--50.7)} \\
\cmidrule{2-6}
 & \multirow{3}{*}{Fact Inversion} 
 & Consistent   & \heatmanual{5}{57.7\% $\pm$ 3.3\% (51.1--64.2)} & \heatmanual{2}{32.3\% $\pm$ 3.1\% (26.6--38.4)} & \heatmanual{2}{32.3\% $\pm$ 3.1\% (26.6--38.4)} \\
 & & Neutral      & \heatmanual{4}{38.0\% $\pm$ 3.2\% (31.9--44.1)} & \heatmanual{2}{30.6\% $\pm$ 3.0\% (24.9--36.7)} & \heatmanual{2}{27.9\% $\pm$ 3.0\% (22.3--34.1)} \\
 & & Inconsistent & \heatmanual{2}{29.2\% $\pm$ 3.0\% (23.6--35.4)} & \heatmanual{2}{31.0\% $\pm$ 3.0\% (25.3--37.1)} & \heatmanual{2}{27.1\% $\pm$ 2.9\% (21.4--33.2)} \\
\cmidrule{2-6}
 & \multirow{3}{*}{FSAP-InterQ} 
 & Consistent   & \heatmanual{2}{32.7\% $\pm$ 3.1\% (26.6--38.9)} & \heatmanual{1}{16.6\% $\pm$ 2.5\% (11.8--21.4)} & \heatmanual{1}{17.5\% $\pm$ 2.5\% (12.7--22.7)} \\
 & & Neutral      & \heatmanual{1}{21.0\% $\pm$ 2.7\% (15.7--26.6)} & \heatmanual{1}{18.3\% $\pm$ 2.5\% (13.5--23.6)} & \heatmanual{1}{15.3\% $\pm$ 2.4\% (10.9--20.1)} \\
 & & Inconsistent & \heatmanual{1}{20.1\% $\pm$ 2.6\% (15.3--25.3)} & \heatmanual{1}{17.5\% $\pm$ 2.5\% (12.7--22.7)} & \heatmanual{1}{14.8\% $\pm$ 2.4\% (10.5--19.7)} \\
\cmidrule{2-6}
 & \multirow{3}{*}{FSAP-IntraQ} 
 & Consistent   & \heatmanual{1}{17.5\% $\pm$ 2.5\% (12.7--22.7)} & \heatmanual{0}{2.6\% $\pm$ 1.1\% (0.9--4.8)} & \heatmanual{0}{1.3\% $\pm$ 0.8\% (0.0--3.1)} \\
 & & Neutral      & \heatmanual{0}{6.1\% $\pm$ 1.6\% (3.1--9.2)} & \heatmanual{0}{4.4\% $\pm$ 1.4\% (1.7--7.0)} & \heatmanual{0}{2.2\% $\pm$ 1.0\% (0.4--4.4)} \\
 & & Inconsistent & \heatmanual{0}{3.5\% $\pm$ 1.2\% (1.3--6.1)} & \heatmanual{0}{2.2\% $\pm$ 1.0\% (0.4--4.4)} & \heatmanual{0}{{0.0\% $\pm$ 0.0\% (0.0--0.0)}} \\
\cmidrule{2-6}
 & \multirow{3}{*}{Liar} 
 & Consistent   & \heatmanual{1}{{13.1\% $\pm$ 2.2\% (8.7--17.9)}} & \heatmanual{0}{{1.7\% $\pm$ 0.9\% (0.4--3.5)}} & \heatmanual{0}{{0.0\% $\pm$ 0.0\% (0.0--0.0)}} \\
 & & Neutral      & \heatmanual{0}{{1.3\% $\pm$ 0.8\% (0.0--3.1)}} & \heatmanual{0}{{2.2\% $\pm$ 1.0\% (0.4--4.4)}} & \heatmanual{0}{{0.4\% $\pm$ 0.4\% (0.0--1.3)}} \\
 & & Inconsistent & \heatmanual{0}{{3.9\% $\pm$ 1.3\% (1.7--6.6)}} & \heatmanual{0}{{0.0\% $\pm$ 0.0\% (0.0--0.0)}} & \heatmanual{0}{0.4\% $\pm$ 0.4\% (0.0--1.3)} \\
\bottomrule
\end{tabular}
}
\vspace{-1em}
\end{table}

\begin{table}[t]
\centering
\caption{Results of \textbf{single-document setup} (\%, mean $\pm$ SD with 95\% CI) of ground-truth alignment rate for \textbf{open-source LLMs} on \textbf{TREC 2020}. Providing helpful documents yields the highest ground-truth alignment rates across all models, while the Liar documents result in the lowest. The pattern of results is consistent across all three models, with cells shaded by mean value to indicate relative performance (darker shading corresponds to higher alignment rate).}
\label{tab:stance_open_llm_2020}
\renewcommand{\arraystretch}{1.2}
\setlength{\tabcolsep}{6pt}
\resizebox{\textwidth}{!}{%
\begin{tabular}{lllccc}
\toprule
\textbf{Setting} & \textbf{Context} & \textbf{Query Type} & \textbf{DeepSeek-R1-Distill-Qwen-32B} & \textbf{Phi-4} & \textbf{LLaMA-3 8B Instruct} \\
\midrule
\multirow{3}{*}{Non-RAG} & – 
 & Consistent   & \heatmanual{9}{95.3\% $\pm$ 4.4\% (86.4--100.0)} & \heatmanual{9}{{95.3\% $\pm$ 4.4\% (86.4--100.0)}} & \heatmanual{9}{90.8\% $\pm$ 6.1\% (77.3--100.0)} \\
 & & Neutral      & \heatmanual{8}{86.3\% $\pm$ 7.3\% (72.7--100.0)} & \heatmanual{9}{{95.3\% $\pm$ 4.4\% (86.4--100.0)}} & \heatmanual{9}{90.8\% $\pm$ 6.1\% (77.3--100.0)} \\
 & & Inconsistent & \heatmanual{9}{95.3\% $\pm$ 4.4\% (86.4--100.0)} & \heatmanual{8}{86.3\% $\pm$ 7.3\% (72.7--100.0)} & \heatmanual{8}{86.3\% $\pm$ 7.3\% (72.7--100.0)} \\
\midrule
 & \multicolumn{5}{l}{\textbf{Original Documents}} \\
\cmidrule{2-6}
\multirow{24}{*}{RAG} & \multirow{3}{*}{Helpful} 
 & Consistent   & \heatmanual{9}{{95.8\% $\pm$ 1.4\% (92.7--98.4)}} & \heatmanual{9}{91.7\% $\pm$ 2.0\% (87.6--95.3)} & \heatmanual{9}{{92.3\% $\pm$ 1.9\% (88.1--95.9)}} \\
 & & Neutral      & \heatmanual{9}{{92.8\% $\pm$ 1.9\% (89.1--96.4)}} & \heatmanual{9}{93.3\% $\pm$ 1.8\% (89.6--96.4)} & \heatmanual{9}{{91.2\% $\pm$ 2.0\% (87.0--94.8)}} \\
 & & Inconsistent & \heatmanual{9}{{95.3\% $\pm$ 1.5\% (92.2--97.9)}} & \heatmanual{9}{{93.8\% $\pm$ 1.7\% (90.2--96.9)}} & \heatmanual{9}{{92.8\% $\pm$ 1.9\% (89.1--96.4)}} \\
\cmidrule{2-6}
 & \multirow{3}{*}{Harmful} 
 & Consistent   & \heatmanual{8}{84.4\% $\pm$ 3.3\% (77.7--90.9)} & \heatmanual{7}{72.8\% $\pm$ 4.1\% (64.5--80.2)} & \heatmanual{6}{62.9\% $\pm$ 4.4\% (53.0--71.1)} \\
 & & Neutral      & \heatmanual{6}{62.0\% $\pm$ 4.4\% (52.9--70.2)} & \heatmanual{6}{63.7\% $\pm$ 4.4\% (54.5--71.9)} & \heatmanual{5}{57.9\% $\pm$ 4.5\% (48.8--66.9)} \\
 & & Inconsistent & \heatmanual{6}{64.5\% $\pm$ 4.4\% (56.2--72.7)} & \heatmanual{5}{57.9\% $\pm$ 4.5\% (48.8--66.9)} & \heatmanual{6}{61.2\% $\pm$ 4.5\% (52.9--69.4)} \\
\cmidrule{2-6}
 & \multicolumn{5}{l}{\textbf{Adversarial Documents}} \\
\cmidrule{2-6}
 & \multirow{3}{*}{Rewriter} 
 & Consistent   & \heatmanual{9}{96.4\% $\pm$ 1.4\% (93.3--98.4)} & \heatmanual{8}{80.3\% $\pm$ 2.9\% (74.6--86.0)} & \heatmanual{7}{77.8\% $\pm$ 3.0\% (71.5--83.4)} \\
 & & Neutral      & \heatmanual{7}{75.2\% $\pm$ 3.1\% (68.9--81.3)} & \heatmanual{7}{77.8\% $\pm$ 3.0\% (71.5--83.4)} & \heatmanual{7}{68.5\% $\pm$ 3.4\% (61.7--75.1)} \\
 & & Inconsistent & \heatmanual{7}{72.1\% $\pm$ 3.3\% (65.8--78.2)} & \heatmanual{7}{75.7\% $\pm$ 3.1\% (69.4--81.9)} & \heatmanual{7}{79.3\% $\pm$ 2.9\% (73.6--85.0)} \\
\cmidrule{2-6}
 & \multirow{3}{*}{Paraphraser} 
 & Consistent   & \heatmanual{9}{95.3\% $\pm$ 1.5\% (92.2--97.9)} & \heatmanual{7}{80.9\% $\pm$ 2.9\% (75.1--86.0)} & \heatmanual{7}{72.1\% $\pm$ 3.3\% (65.8--78.2)} \\
 & & Neutral      & \heatmanual{7}{73.1\% $\pm$ 3.2\% (66.8--79.3)} & \heatmanual{7}{71.0\% $\pm$ 3.3\% (64.2--77.2)} & \heatmanual{6}{65.9\% $\pm$ 3.4\% (59.1--72.5)} \\
 & & Inconsistent & \heatmanual{7}{74.7\% $\pm$ 3.2\% (68.4--80.8)} & \heatmanual{7}{71.6\% $\pm$ 3.3\% (65.3--77.7)} & \heatmanual{7}{72.6\% $\pm$ 3.2\% (66.3--78.8)} \\
\cmidrule{2-6}
 & \multirow{3}{*}{Fact Inversion} 
 & Consistent   & \heatmanual{9}{92.8\% $\pm$ 1.9\% (89.1--96.4)} & \heatmanual{7}{76.2\% $\pm$ 3.1\% (69.9--82.4)} & \heatmanual{7}{74.1\% $\pm$ 3.2\% (67.9--80.3)} \\
 & & Neutral      & \heatmanual{7}{74.1\% $\pm$ 3.2\% (67.9--80.3)} & \heatmanual{7}{74.1\% $\pm$ 3.2\% (67.9--80.3)} & \heatmanual{6}{64.3\% $\pm$ 3.5\% (57.5--71.0)} \\
 & & Inconsistent & \heatmanual{7}{73.6\% $\pm$ 3.2\% (67.4--79.8)} & \heatmanual{7}{69.0\% $\pm$ 3.4\% (62.2--75.6)} & \heatmanual{6}{65.9\% $\pm$ 3.4\% (59.1--72.5)} \\
\cmidrule{2-6}
 & \multirow{3}{*}{FSAP-InterQ} 
 & Consistent   & \heatmanual{7}{79.3\% $\pm$ 2.9\% (73.6--85.0)} & \heatmanual{5}{51.3\% $\pm$ 3.6\% (44.6--58.5)} & \heatmanual{4}{46.6\% $\pm$ 3.6\% (39.9--53.9)} \\
 & & Neutral      & \heatmanual{4}{49.8\% $\pm$ 3.6\% (43.0--57.0)} & \heatmanual{4}{45.6\% $\pm$ 3.6\% (38.3--52.8)} & \heatmanual{4}{42.0\% $\pm$ 3.6\% (35.2--49.2)} \\
 & & Inconsistent & \heatmanual{5}{52.3\% $\pm$ 3.6\% (45.1--59.1)} & \heatmanual{4}{43.0\% $\pm$ 3.6\% (36.3--50.3)} & \heatmanual{4}{40.4\% $\pm$ 3.5\% (33.7--47.7)} \\
\cmidrule{2-6}
 & \multirow{3}{*}{FSAP-IntraQ} 
 & Consistent   & \heatmanual{6}{66.4\% $\pm$ 3.4\% (59.6--73.1)} & \heatmanual{2}{32.6\% $\pm$ 3.4\% (25.9--39.4)} & \heatmanual{2}{33.2\% $\pm$ 3.4\% (26.4--39.9)} \\
 & & Neutral      & \heatmanual{2}{28.5\% $\pm$ 3.2\% (22.3--35.2)} & \heatmanual{2}{26.4\% $\pm$ 3.1\% (20.2--32.6)} & \heatmanual{2}{23.8\% $\pm$ 3.0\% (18.1--30.1)} \\
 & & Inconsistent & \heatmanual{2}{28.5\% $\pm$ 3.2\% (22.3--35.2)} & \heatmanual{2}{23.8\% $\pm$ 3.0\% (18.1--30.1)} & \heatmanual{1}{20.7\% $\pm$ 2.9\% (15.5--26.4)} \\
\cmidrule{2-6}
 & \multirow{3}{*}{Liar} 
 & Consistent   & \heatmanual{5}{{51.8\% $\pm$ 3.6\% (44.6--59.1)}} & \heatmanual{1}{{17.1\% $\pm$ 2.7\% (11.9--22.8)}} & \heatmanual{1}{{19.7\% $\pm$ 2.8\% (14.5--25.4)}} \\
 & & Neutral      & \heatmanual{2}{{23.3\% $\pm$ 3.0\% (17.6--29.5)}} & \heatmanual{1}{{14.5\% $\pm$ 2.5\% (9.8--19.7)}} & \heatmanual{1}{{19.2\% $\pm$ 2.8\% (14.0--24.9)}} \\
 & & Inconsistent & \heatmanual{2}{{22.8\% $\pm$ 3.0\% (17.1--29.0)}} & \heatmanual{1}{{11.9\% $\pm$ 2.3\% (7.8--16.6)}} & \heatmanual{1}{{12.5\% $\pm$ 2.4\% (7.8--17.1)}} \\
\bottomrule
\end{tabular}
}
\vspace{-1em}
\end{table}

To determine the impact of retrieved evidence on LLM behavior, we compare ground-truth alignment rate in the Non-RAG baseline, which relies solely on the prompt, with the single-document setup, where a single document serves as context. 
Figure~\ref{fig:trec-two-years-gpt41} illustrates these results for GPT-4.1 on TREC 2020 and TREC 2021, while Tables~\ref{tab:stance-closed-llms_2021}--\ref{tab:stance_open_llm_2020} report ground-truth alignment rates across all closed- and open-source models, including the mean, standard deviation (SD), and 95\% confidence intervals (CI) obtained via bootstrapping. Since it is not feasible to include plots for every model in the paper, we provide the corresponding visualizations in the project’s github repository \footnote{\url{https://github.com/shakibaam/RAG_ROBUSTNESS_EVAL/tree/main/Plots/gemini2.0flash/Single-Document}}, while Tables~\ref{tab:stance-closed-llms_2021}--\ref{tab:stance_open_llm_2020} present the key comparative results for all models used in the single-document experiment.

In both the TREC 2020 and TREC 2021 datasets, adding helpful retrieved context causes clear improvements in ground-truth alignment rate compared to the Non-RAG baseline, with a single supporting document markedly improving alignment rate. For instance, as shown in Table~\ref{tab:stance-closed-llms_2021}, when GPT-4.1 is evaluated on TREC 2021 and paired with a helpful document, its alignment rate goes from 88.9\% (neutral user query, Non-RAG baseline) to 98.2\%. Inconsistent user queries also show similar improvements, with alignment rate going from 66.7\% (Non-RAG baseline) to 93.0\% with helpful evidence, and comparable trends are observed across other models on TREC 2021 and TREC 2020. For example, GPT-5 evaluated on TREC 2020, as reported in Table~\ref{tab:stance_closed_llm_2020}, shows an increase from 86.3\% in the Non-RAG baseline with a neutral user query to 94.8\% with helpful context, while DeepSeek-R1-Distill-Qwen-32B improves from 86.3\% to 92.8\% under the same user query condition on TREC 2020 with results reported in Table~\ref{tab:stance_open_llm_2020}.

Conversely, in both TREC 2020 and TREC 2021, harmful documents consistently degrade performance. For GPT-4.1 on TREC 2021 (Table~\ref{tab:stance-closed-llms_2021}), the alignment rate falls to 37.7\% with neutral user queries and 32.6\% with inconsistent user queries, while for GPT-5 drops to 43.4\% and 36.5\%, respectively. These results demonstrate that retrieved documents actively influence model alignment, with the ability to markedly enhance or diminish it rather than merely supplying contextual information. The consistency of this pattern across both closed- and open-source models underscores the central role of context in shaping LLM behavior when compared to the Non-RAG baseline, and this finding addresses RQ1 by showing that context definitely impacts LLM responses.

\subsection{RQ2: To what extent does the type of retrieved context influence robustness?}
\label{sec:rq2-results}

With RQ1 confirming that context matters, RQ2 examines how its type, helpful, harmful, or adversarial, affects ground-truth alignment. Results from the single-document setup (Tables~\ref{tab:stance-closed-llms_2021}--\ref{tab:stance_open_llm_2020}) reveal a consistent pattern across both models and datasets, as helpful documents raise alignment rate substantially above the Non-RAG baseline, harmful documents reduce it, and adversarial documents often cause robustness to collapse almost entirely.

For closed-source models on TREC 2021 (Table~\ref{tab:stance-closed-llms_2021}), 
helpful documents push alignment rate close to the ceiling for almost all user query framings. For instance, ground-truth alignment rate rises from 88.9\% in Non-RAG baseline to 98.2\% for GPT-4.1 for neutral user queries, from 81.6\% to 98.7\% for GPT-5, and from 70.4\% to 97.4\% for Claude-3.5-Haiku. Under the same neutral setting on TREC 2021, harmful evidence reduces these gains and pushes alignment rate down to 37.7\% for GPT-4.1, 43.4\% for GPT-5, and 36.0\% for Claude-3.5-Haiku. Furthermore, alignment rate is nearly eliminated by strong adversarial documents such as Liar, with GPT-4.1 dropping to 4.4\%, GPT-5 to 2.6\%, and Claude-3.5-Haiku to 0.9\% for the same neutral condition.

Open-source models follow the same pattern. As shown in Table~\ref{tab:stance_open_llm_2021} for TREC 2021, the alignment rate of DeepSeek-R1-Distill-Qwen-32B improves from 63.0\% in the Non-RAG baseline to 92.6\% with helpful documents for neutral user queries, but collapses to just 1.3\% under Liar inputs. A similar trend appears for Phi-4, which rises from 66.7\% in the Non-RAG baseline to 96.5\% with helpful evidence, but drops sharply to 2.2\% with Liar documents. LLaMA-3 8B Instruct shows the same pattern, moving from 81.6\% without context to 93.9\% with helpful inputs, before falling to just 0.4\% under Liar documents. Even weaker adversarial documents, such as Rewriter and Paraphraser, substantially reduce alignment rate, often into the 55–70\% range across all open-source models. The same pattern is observed for TREC 2020 as well, but with consistently higher rates, as detailed in the following paragraphs.
We emphasize neutral user query results here because RQ1–RQ3 focus primarily on the effect of retrieved context, while the role of query framing is analyzed separately in Section~\ref{sec:rq4-results}.

Across both datasets and model families, the relative effectiveness of adversarial strategies follows a consistent ordering: \textit{Liar} $\gg$ \textit{FSAP-IntraQ} $>$ \textit{FSAP-InterQ} $>$ \textit{Fact Inversion} $>$ \textit{Paraphraser} $\approx$ \textit{Rewriter}. This hierarchy is a result of how adversarial documents are made. Liar documents are constructed from scratch based on the query and an incorrect stance, so they are not constrained to the wording or rhetorical framing of pre-existing harmful content. This independence enables more fluent phrasing and broader topic coverage, making misinformation more confident and persuasive, and thus more challenging for LLMs. In contrast, Paraphraser and Rewriter documents inherit limitations from pre-existing harmful material, which can make them appear less natural.

In cross-model comparison, GPT-4.1 and GPT-5 achieve the highest levels of alignment rate when paired with helpful documents; however, both demonstrate comparable vulnerabilities when confronted with adversarial evidence. Claude-3.5-Haiku has significant performance drops in the presence of harmful or adversarial content; on the other hand, DeepSeek-R1-Distill-Qwen-32B performs well and often obtains levels that are similar to closed-source models when provided with helpful documents. Phi-4 and LLaMA-3 8B Instruct achieves a lower overall alignment rate but follows the same qualitative trends. Since these patterns are consistent across all models, we selected GPT-4.1 as the representative model for subsequent experiments due to its relatively high performance and speed. GPT-5 was not released at the time these experiments were first conducted and was included afterwards.

Another important finding of this experiment is the difference at the dataset level, where models consistently display a higher alignment rate on TREC 2020 (COVID-19 misinformation) than on TREC 2021 (general health misinformation). For example, TREC 2021 includes queries about misinformation on everyday remedies and niche treatments, such as using duct tape for warts and toothpaste for pimples, which rarely reach the same level of wide public discourse, although the ground truth for these questions is based on evidence. This underscores an unexpected outcome from our research, suggesting that LLMs appear considerably more robust against adversarial manipulation in the COVID-19 context. This difference may be due to post-training alignment placing greater emphasis on topics like COVID-19, which was the subject of widespread misinformation and public attention, while giving limited focus to less prominent areas of health.

Overall, our results address RQ2 by showing that the type and quality of retrieved context substantially affect LLM robustness, by demonstrating that helpful documents consistently improve alignment rate relative to the baseline, whereas harmful and adversarial documents undermine stability and reliability, in some cases reducing robustness to the point of complete failure.

While it may appear self-evident that helpful evidence would improve alignment rate and harmful or adversarial evidence would reduce it, our results move beyond speculation by empirically quantifying both the magnitude and the consistency of these effects across datasets and model families. The measurements establish concrete performance bounds. Alignment rate can approach near-maximum levels by providing a single helpful document, yet it can collapse to almost zero under adversarial documents such as Liar. As outlined in Section~\ref{sec:introduction}, this systematic analysis offers the empirical evidence required to turn abstract concerns about adversarial susceptibility into evidence-based design principles for RAG systems. In line with RQ1, the findings highlight that context can significantly shape outcomes and must be safeguarded against harmful or adversarial inputs.

\subsection{RQ3: In what ways do orders and combinations of multiple evidence sources shape model behavior?}
\label{sec:rq3-results}

Findings from RQ1 and RQ2 show that context type strongly shapes ground-truth alignment, which motivates an assessment of more complex scenarios with multiple documents presented in different orders and compositions, leading to RQ3. To address this question, we examine how models respond when exposed to multiple sources of evidence, focusing on two key dimensions: (i) the \emph{order} in which helpful and misleading documents are presented, and (ii) the \emph{composition} of the retrieval pool when evidence is naturally or systematically skewed. 
We investigate these dimensions through three complementary setups: the paired-document setup, the passage-based pooling setup, and the bias-controlled pooling setup, which are analyzed in the following sections.

\subsubsection{Paired-Document Analysis}
\label{sec:paired-docs-results}

\begin{figure}[t]
  \centering
  \includegraphics[width=\linewidth]{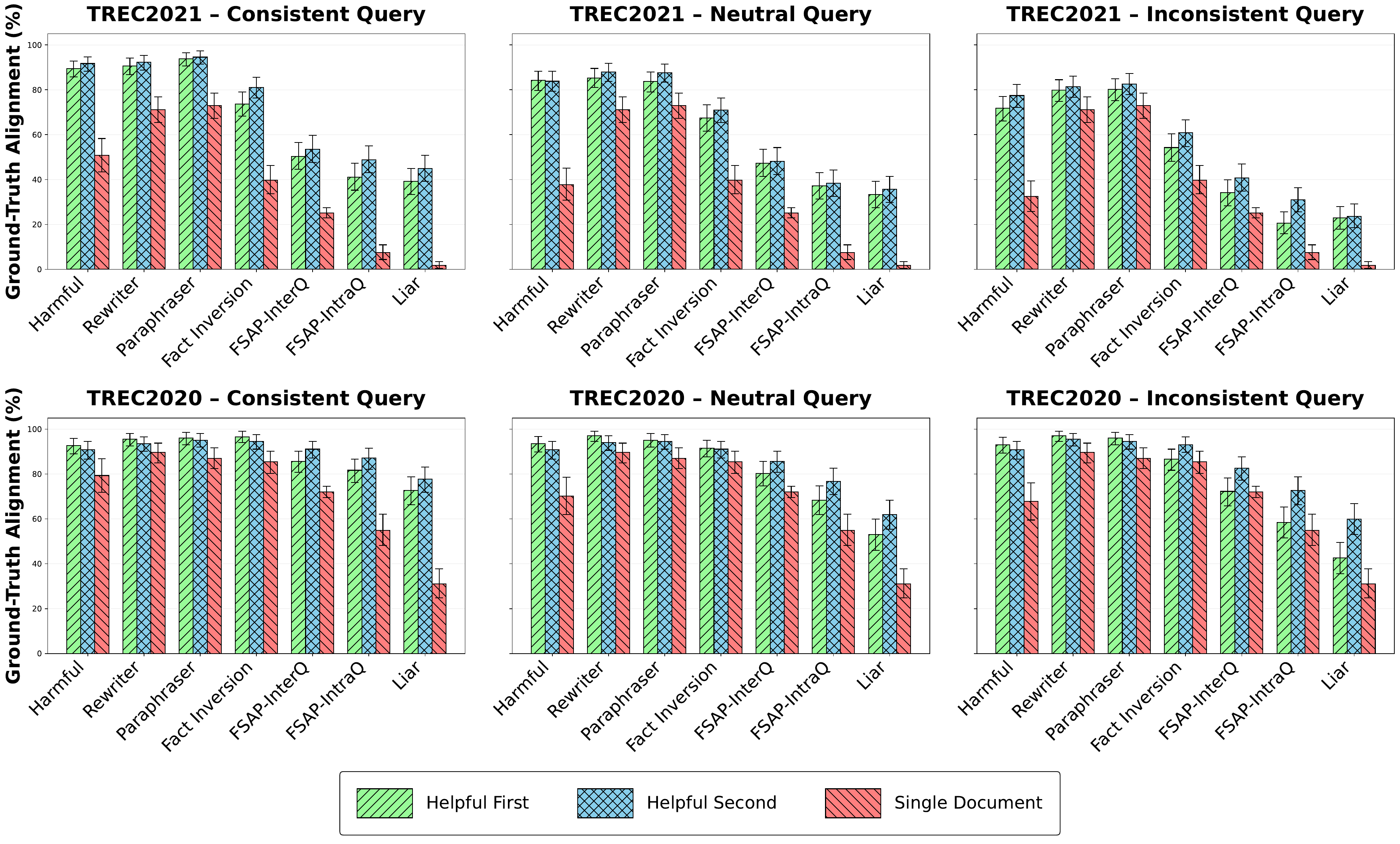}
  \caption{Results of \textbf{paired-document setup} for \textbf{GPT-4.1} on \textbf{TREC 2020} and \textbf{TREC 2021} with consistent, neutral, and inconsistent user queries. The x-axis shows the document type with which the helpful document is paired (either first or second), and the y-axis reports ground-truth alignment rate (\%). Error bars indicate 95\% confidence intervals estimated via bootstrapping. Bars compare helpful-first, helpful-second, and single-document inputs (harmful/adversarial only). Helpful placement shows only minor differences, while single-document inputs yield substantially lower ground-truth alignment rates.}
  \label{fig:paired_result}
\end{figure}

Figure~\ref{fig:paired_result} shows ground-truth alignment rate when a helpful document is either first (green, ``Helpful First'') or second (blue, ``Helpful Second'') relative to a harmful/adversarial document, compared with the single-document baseline (red). The results show that the order of the documents has minimal impact on ground-truth alignment. The Helpful First and Helpful Second conditions deliver nearly the same results, with confidence intervals that mostly overlap. For instance, on TREC 2021 under the consistent user query, the alignment rate of LLM responses rises from only 1.7\% (CI 0.4--3.5) in the single-document baseline to 39.1\% (CI 33.3--45.0) when helpful evidence appears first and 45.0\% (CI 39.1--50.8) when it appears second for Liar inputs. Similarly, for FSAP-IntraQ documents, alignment rate improves from 7.4\% (CI 4.4--10.9) to 41.1\% (CI 35.3--47.3) when helpful content precedes and 48.8\% (CI 43.0--55.0) when it follows the adversarial document. On TREC 2020, robustness is consistently higher. For example, under the neutral user query, for FSAP-IntraQ documents, alignment rate rises from 54.9\% (CI 48.2--62.2) to 68.3\% (CI 61.9--74.8) when helpful material comes first and 76.7\% (CI 70.8--82.7) when it is placed second relative to them. These results show that presenting helpful evidence, regardless of position, minimizes the impact of harmful or adversarial content. Additionally, discrepancies among datasets persist, as TREC 2020 (COVID-19) queries demonstrate greater robustness than the more general health queries in TREC 2021, thereby reinforcing the previously identified dataset-level variations.

\subsubsection{Passage-Based Pooling Analysis} 
\label{sec:natural-pool-performance} 

In the passage-based pooling setup, the MonoT5 reranker provides the top-10 passages for each query, obtained by sliding a 512-word window with 256-word overlap across documents and ranking the resulting segments. Since the adversarial documents are designed to rank highly, these pools are heavily skewed toward this adversarial content, with more than nine of the top-10 segments being adversarial. For TREC 2020, the top-10 pools comprise $\sim$92\% adversarial, 5\% helpful, and 3\% harmful passages, whereas for TREC 2021 the distribution is $\sim$94\% adversarial, 5\% helpful, and 2\% harmful passages.
As shown in Fig.~\ref{fig:pool_result}, the dominance of adversarial segments in the passage-based pool leads to consistently low alignment rates. On TREC 2020, GPT-4.1 reaches 43.7\% under consistent user queries, but performance drops to 24.9\% with neutral user queries and remains low at 24.9\% under inconsistent user queries. The degradation is even worse on TREC 2021, where the alignment rate drops to 17.3\% with consistent user queries, 13.0\% with neutral user queries, and only 4.3\% with inconsistent user queries. These findings emphasize retrieval as the primary bottleneck in RAG pipelines and demonstrate that the composition of the evidence pool can significantly undermine robustness, irrespective of user query framing.

\subsubsection{Bias-Controlled Pooling Analysis}
\label{sec:biased-pool-performance}

\begin{figure}[t]
  \centering
  \includegraphics[width=\linewidth]{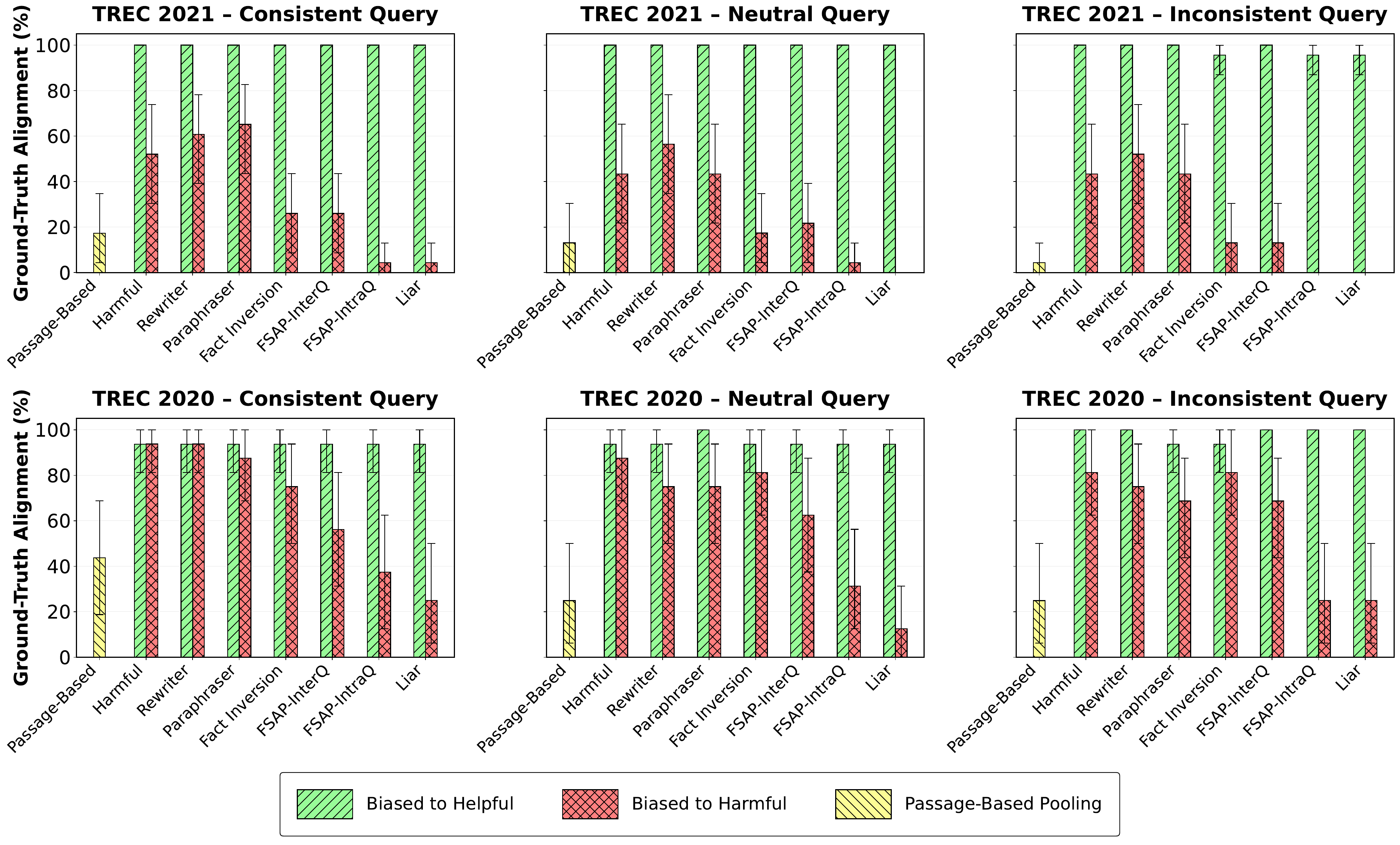}
  \caption{Results of \textbf{passage-based pooling} and \textbf{biased-controlled pooling setups} for \textbf{GPT-4.1} on \textbf{TREC 2020} and \textbf{TREC 2021} across consistent, neutral, and inconsistent user queries. The x-axis includes the passage-based pooling baseline as well as the different document types used in biased conditions, where helpful documents are paired with harmful or adversarial ones. The y-axis reports ground-truth alignment rate (\%). Error bars indicate 95\% confidence intervals estimated via bootstrapping. Results show helpful-biased pools yield near-perfect alignment, whereas harmful and passage-based pools sharply reduce it.}
  \label{fig:pool_result}
\end{figure}

To examine the impact of systematic skew in retrieval, we construct pools biased toward either helpful or harmful content. In the condition biased toward harmful, the ground-truth alignment rate declines markedly. On TREC 2021, as shown in Fig.~\ref{fig:pool_result}, GPT-4.1 achieves only a 26.04\% alignment rate when the pool is biased toward harmful content with Fact Inversion documents, and 4.35\% when biased with Liar documents, even under consistent user queries. The situation worsens under inconsistent user queries, where providing Liar examples leads to a 0\% alignment rate, and presenting FSAP-IntraQ documents collapses the rate to 0\% as well. By contrast, TREC 2020 demonstrates somewhat greater resilience. Although alignment rate against Rewriter documents remains as high as 93.76\% and Paraphraser at 87.52\% under consistent user queries, it still collapses for the most adversarial harmful cases, dropping to 24.95\% for Liar and 37.4\% for FSAP-IntraQ documents.

In the biased-toward-helpful condition, alignment levels are close to the best possible. On both datasets, consistent and neutral user queries frequently sustain a 100\% alignment rate, with LLM responses across all attack categories in TREC 2021 reaching a full alignment rate when helpful documents are included in the pool. Even inconsistent user queries remain robust: in TREC 2020, pooled conditions containing Rewriter or FSAP-InterQ documents yield a 100\% alignment rate, and Fact Inversion pools reach 95.63\% on TREC 2021. Once helpful content dominates the retrieval pool, the influence of user query framing diminishes, as supportive context outweighs misleading presuppositions.

Overall, the bias-controlled experiments confirm that the composition of retrieved evidence plays a critical role in RAG systems' robustness. Pools dominated by helpful documents consistently lead to high alignment rates, often exceeding 90–100\%, while harmful-biased pools drastically reduce reliability, in some cases collapsing alignment rate to nearly zero. These results underscore that the stability of RAG systems is fundamentally dependent on the distribution of their evidence sources, which highlights the retrieval stage as the key vulnerability in pipeline design.

\subsection{RQ4: How do different query framings affect stance alignment?}
\label{sec:rq4-results}

After discussing the significance of evidence type (RQ2) and evidence order/combination (RQ3), we move on to RQ4, which assesses the impact of user query framing (consistent, neutral, and inconsistent) on the results of ground-truth alignment. Prior work has consistently shown that LLM performance is highly sensitive to prompt formulation. Even minor variations in wording, token placement, or tone can lead to substantially different outcomes, and several studies have proposed methods to stabilize predictions under such alterations~\cite{qiang2024prompt,razavi2025benchmarking,mao2023prompt,wang2024assessing,mizrahi2024state,li2025enhancing,perccin2025investigating,hu2024prompt}. Motivated by these results, we revisit our experimental setups with query sensitivity in mind.

We analyze user query formulation effects across all four experimental setups: the single-document setup, the paired-document setup, the passage-based pooling setup, and the bias-controlled pooling setup. Figures~\ref{fig:trec-two-years-gpt41}--\ref{fig:pool_result} and Tables~\ref{tab:stance-closed-llms_2021}--\ref{tab:stance_open_llm_2020} summarize results across models, datasets, and retrieval conditions.

Framing has a persistent impact on ground-truth alignment rate, and the order of user query types remains the same across models and datasets (Consistent $>$ Neutral $>$ Inconsistent), which directly addresses RQ4 by showing that the framing of user queries impacts ground-truth alignment outcomes. The same hierarchy emerges in the single-document setup (Fig.\ref{fig:trec-two-years-gpt41}), persists when multiple sources of evidence are paired (Fig.\ref{fig:paired_result}), and carries over to more passage-based pooling setups (Fig.~\ref{fig:pool_result}).

Consistent user queries, those which align with the ground-truth stance of the answer, reliably produce the highest alignment rates by framing the model’s reasoning in a direction that agrees with available evidence; on the other hand, inconsistent user queries embed contradictory presuppositions that actively mislead the model, often overwhelming the corrective signal of helpful documents. While explicit stance bias is eliminated by neutral user queries, models are still able to base their responses on the evidence they have retrieved, resulting in intermediate performance. These results show that models reflect the quality of retrieved evidence (RQ2) and systematically inherit bias from how user queries are phrased. User query formulation, therefore, plays a central role in shaping ground-truth alignment outcomes (RQ4).

\section{Concluding Discussion}
\label{sec:discussion}

A central finding of this work is that query framing and retrieval context interact to affect how well LLMs handle health misinformation. The results consistently show that user queries and the type of evidence retrieved have a significant impact on model behavior, with the content distribution of the retrieval pool shaping outcomes alongside the quality of individual documents. In particular, the inclusion of helpful evidence, even in the presence of adversarial or harmful content, can protect against misalignment.

The empirical analysis demonstrates that ground-truth alignment is most severely harmed by Liar and FSAP documents, with effects that are consistent across both open-source and closed-source model families. Our observations are consistent with previous studies \citet{zou2402poisonedrag,BadRAG,chaudhari2024phantom,wang2025derag} showing that RAG systems are sensitive to the information provided as context, and when this context is malicious or adversarially manipulated, models can be misled into producing incorrect or biased output. This reinforces the broader evidence that the reliability of RAG pipelines is dependent not only on retrieval accuracy but also on the integrity of the documents used to ground generation. However, adding helpful documents often makes performance superior to the Non-RAG baseline. Query framing is also a key factor in shaping model behavior, with consistent user queries mostly improving alignment and conflicting queries reducing it.

In addition, our results reveal a consistent robustness gap between COVID-19 queries in TREC 2020 and general health queries in TREC 2021, likely due to extensive exposure to pandemic-related misinformation during training that may have reinforced model defenses. In contrast, broader health topics remain more vulnerable, which suggests a potential imbalance in how current LLMs are safeguarded across medical domains. These outcomes indicate a more extensive asymmetry in alignment practices, wherein robustness is shown to be enhanced in response to prominent crises but is insufficiently applied to other medically significant domains.

Our study provides important evidence on the vulnerabilities and unintended risks associated with generative AI in potentially high-stakes domains such as healthcare. Our findings demonstrate that adversarial documents can significantly disrupt model behavior, undermining the reliability of even state-of-the-art RAG systems. These failures occur regardless of the size or architecture of the model and underscore the retrieval stage as the primary point of vulnerability. Once malicious content is promoted to the top ranks by the retriever or reranker, LLMs rarely recover, even when supported by consistent user queries. This highlights a broader concern: generative AI systems are not only prone to adversarial manipulation but can also amplify harmful content when such information is injected into their context. The results emphasize that enhancing the generation component alone is insufficient to ensure trustworthiness. Robust defenses must also address vulnerabilities in the retrieval pipeline to prevent harmful content from reaching the generation stage.

\section{Limitations}
\label{sec:limitations}
Our study advances the understanding of RAG robustness under adversarial evidence, but it has limitations. First, we rely on ground-truth alignment as the primary evaluation metric, which does not fully capture other important aspects of response quality, such as completeness, clarity, or potential user trust effects. Future work should employ multidimensional evaluations to provide a more comprehensive assessment of system performance in sensitive domains.

Second, our design assumes that relevant documents are already retrieved. This lets us isolate how document content and query framing influence model behavior, but omits retrieval-stage factors such as ranking errors and bias. Future work should extend this analysis to full RAG pipelines to examine how adversarial documents interact with retrievers and rerankers under real-world conditions.

Finally, our evaluation relies on automated stance classification with gemini-2.0-flash, validated using gpt-4o-mini. While prior work shows LLMs can align well with human judgments~\cite{gilardi2023chatgpt,chiang-lee-2023-large}, automated classifiers may miss subtle inaccuracies or nuanced risks. Future work should include human-in-the-loop or multi-annotator studies to strengthen validity in safety-critical domains.


\bibliographystyle{ACM-Reference-Format}
\bibliography{references} 

\end{document}